\documentclass[preprints,article,accept,pdftex,moreauthors]{Definitions/mdpi} 

\usepackage{booktabs}
\usepackage{longtable}
\usepackage{array}
\usepackage[table]{xcolor}

\firstpage{1} 
\pubvolume{1}
\issuenum{1}
\articlenumber{0}
\pubyear{2026}
\copyrightyear{2026}
\datereceived{ } 
\daterevised{ } 
\dateaccepted{ } 
\datepublished{ } 

\Title{Reasoning Across Scales: Learning Challenges and Opportunities for Immersive Technologies in Chip Design Education}

\Author{
Satya Kamala Immidisetty $^{1}$\orcidA{}, 
Dishita Turakhia $^{2}$\orcidB{}
}

\AuthorNames{Satya Kamala Immidisetty, Dishita Turakhia}

\address{%
$^{1}$ \quad Department of Computer Science, Emory University; satya.kamala.immidisetty@emory.edu\\
$^{2}$ \quad Department of Computer Science, Emory University; dishita.turakhia@emory.edu
}

\abstract{
As the demand for semiconductor talent grows, there is an increasing need for educational approaches that prepare learners for the complexity of modern chip design. Current instruction relies largely on lectures, textbooks, software exercises, and limited laboratory experiences. While these methods provide essential foundations, they offer limited support for understanding the spatial, multi-scale, and interconnected nature of chip design. Immersive technologies, particularly virtual reality (VR), offer opportunities to visualize and interact with complex structures and processes, yet where and how they can meaningfully support semiconductor education remains underexplored.
To investigate this gap, we combined chip design curriculum analysis with semi-structured interviews involving five participants with instructional and learner perspectives. Our analysis characterized four categories of learning demands: descriptive knowledge, procedural knowledge, relational and design-focused reasoning, and strategic decision-making. We also identified recurring challenges in connecting abstraction levels, interpreting spatial structures and temporal behavior, bridging theory and practice, and reasoning about design trade-offs and downstream consequences. Drawing on these findings and prior research, we developed a conceptual mapping between learning demands and six immersive affordances: spatial visualization, multi-scale navigation, interactive manipulation, temporal dynamics visualization, immediate feedback, and collaboration.
Our findings motivate Cross-Scale Spatial Reasoning (CSSR) as a conceptual lens for understanding how learners connect physical structures, design representations, and engineering consequences across the chip design pipeline. This work contributes an empirically informed characterization of learning challenges, a structured mapping of learning demands to immersive affordances, and concrete design opportunities for immersive semiconductor education. Together, these contributions provide a learning-centered foundation for developing and evaluating immersive environments that support integrated reasoning across scales, representations, and design processes.
}


\begin{document}




\section{Introduction}

The semiconductor industry is projected to exceed USD 1 trillion in annual revenue by 2030~\citep{ref-url1-deloitte}, driven by advances in generative artificial intelligence, large-scale data centers, edge computing, and heterogeneous integration. This growth is accompanied by rising demand for skilled talent, with global projections estimating a need for more than one million additional semiconductor workers by 2030~\citep{ref-SEMI-global-talent-gap,ref-url3-talent-gap}. An aging workforce further intensifies this challenge, with approximately one-third of semiconductor professionals in Western economies approaching retirement~\citep{ref-url3-talent-gap}. Addressing this workforce gap requires not only expanding access to semiconductor education, but also strengthening how students develop the knowledge and skills needed for modern chip design.

Efforts to strengthen the semiconductor education pipeline have emphasized earlier exposure to hardware concepts, hands-on training, and closer alignment between academic instruction and industry practice~\citep{ref-journal4-chips-act}. Yet preparing students for modern chip design involves more than increasing participation or practical exposure. Chip design spans multiple physical scales and levels of abstraction, from semiconductor physics and transistor behavior to digital logic, system architectures, physical implementation, and manufacturing. Learning to navigate this pipeline requires students to understand individual concepts and workflows while also recognizing how decisions at one level shape the behavior, constraints, and performance of the larger system. Developing this integrated understanding calls for educational approaches that connect theoretical knowledge with practical experience~\citep{dewey1938experience,kolb1984experiential}.

Conventional chip design instruction combines lectures, textbooks, software tutorials, and laboratory exercises, providing essential conceptual and technical foundations. However, the relationships students must understand are often distributed across separate diagrams, schematics, code, simulation outputs, and physical layouts. For instance, understanding how RTL descriptions become gate-level circuits, how interconnects traverse multiple physical layers, or how placement and routing decisions affect timing and power requires learners to connect representations across different stages of the design process. When these relationships are not made explicit, students must mentally reconstruct spatial structures, interpret temporal behavior, and trace the downstream consequences of design decisions. Prior work in engineering education has highlighted challenges in teaching complex chip design workflows~\citep{ref-journal7,ref-journal9-razavi2024education}. Although hands-on approaches such as FPGA-based laboratories provide opportunities for active learning~\citep{ref-journal10-gulgonul2025sparking}, there remains a need for instructional support that helps learners reason across spatial, temporal, and system-level relationships.

Immersive technologies, including Virtual Reality (VR) and Augmented Reality (AR), offer a potential way to make these relationships more accessible through interactive visualization and exploration. By bringing spatial structures, dynamic processes, and design consequences into a shared interactive environment, immersive technologies may help learners connect concepts that are otherwise encountered through separate representations. Prior research has explored their potential to support spatial understanding, conceptual learning, and engagement~\citep{dalgarno2010learning,merchant2014effectiveness,radianti2020review}. Applications in biology, chemistry, and physics have used immersive environments to visualize molecular structures and physical phenomena~\citep{ref-proceeding11-dunn2023hands,ref-proceeding12-amaral2024interactive,ref-journal13-strzys2018physics}, while studies in engineering education have examined their role in understanding three-dimensional structures and spatial relationships~\citep{ref-journal15-ka2025analysis}. Together, this research suggests opportunities to explore how immersive interaction might support the interconnected reasoning required in chip design.

Despite this potential, it remains unclear where and how immersive technologies can meaningfully contribute to chip design education. Learning demands vary considerably across the design pipeline, from understanding device behavior and architectural trade-offs to interpreting multilayer layouts, tracing timing dependencies, and diagnosing design violations. These tasks involve different forms of reasoning and may not benefit equally from the same visualization or interaction techniques. Although prior work has explored VR for introductory digital circuit learning~\citep{ref-journal16-paszkiewicz2021vr}, and research in related STEM domains has examined the educational possibilities of immersive technologies~\citep{ref-journal17-vidak2024augmented}, there is limited understanding of how specific immersive capabilities align with the learning demands encountered throughout chip design. Addressing this gap requires moving beyond individual technology demonstrations toward a systematic examination of the educational challenges that immersive environments might help address.

To investigate this gap, we conducted a qualitative study combining chip design curriculum analysis with semi-structured interviews involving five participants with instructional and learner perspectives. We examined the knowledge and reasoning required across the curriculum, alongside the difficulties learners encounter when connecting concepts, representations, and design processes. Through thematic analysis~\citep{ref-journal19-naeem2023step}, we characterized four categories of learning demands: descriptive knowledge, procedural knowledge, relational and design-focused reasoning, and strategic decision-making. Drawing on these findings and prior research in immersive learning~\citep{ref-journal23-parong2018learning,ref-journal24-bailenson2008use,ref-journal25-ragan2010effects}, we then developed a conceptual mapping between these demands and six immersive affordances: spatial visualization, multi-scale navigation, interactive manipulation, temporal dynamics visualization, immediate feedback, and collaboration.

This work makes three contributions: (i) an empirically informed characterization of learning demands and challenges across the chip design curriculum, highlighting the importance of reasoning across physical scales, representations, and interconnected design processes; (ii) a structured mapping between these demands and immersive affordances, identifying where and how immersive technologies may provide educational support; and (iii) concrete design opportunities and practical considerations for developing immersive learning environments for chip design education. Together, these contributions establish a learning-centered foundation for investigating how immersive technologies can complement existing instruction and support the spatial, temporal, and systems-level reasoning involved in chip design.

\section{Related Work}

We review prior work in chip design education, emerging learning technologies, and immersive STEM learning, with particular attention to instructional practices, hands-on design workflows, and VR applications in digital circuit education. This review highlights the need to better understand how immersive technologies can support the broader reasoning demands of the chip design pipeline.

\subsection{Chip Design Curriculum and Instructional Practices}

Chip design education spans several levels of abstraction, from digital and analog electronics to Very Large Scale Integration (VLSI), computer architecture, and embedded systems. Courses typically introduce circuit fundamentals before progressing to more complex workflows involving Register Transfer Level (RTL) design, functional verification, logic synthesis, physical implementation, timing closure, and, in some cases, fabrication and testing~\citep{ref-url26-ucscsiliconvalleyext}. Across this progression, students must learn not only the underlying concepts and procedures but also how logical descriptions translate into physical structures and how design decisions influence system-level behavior.

Instruction traditionally combines lectures, textbooks, circuit diagrams, simulation tools, and laboratory exercises, providing foundations for both conceptual understanding and practical experience. Online courses and Massive Open Online Courses (MOOCs) further extend access through tutorials, demonstrations, and self-paced activities~\citep{ref-journal27-abhishek2023effectiveness}. Yet the knowledge students encounter is often distributed across distinct representations, including HDL code, schematics, timing diagrams, simulation reports, and physical layouts. As research on learning with multiple representations suggests, understanding depends not only on interpreting each representation but also on recognizing the relationships between them~\citep{ainsworth2006deft}. In chip design, this requires learners to connect physical structures, temporal behavior, and implementation constraints across different stages of the design process.

Recent initiatives have sought to make these connections more accessible through hands-on, open-source chip design. Programs such as ZeroToASIC, Tiny Tapeout, and Croc expose learners to end-to-end workflows, from RTL development and verification to physical implementation and tapeout~\citep{zerotoasiccourse,venn2024tinytapeout,zelioli2026croc}. By reducing barriers associated with proprietary Electronic Design Automation (EDA) tools and fabrication, these initiatives create opportunities for students to engage with authentic design tasks. Tiny Tapeout, for example, provides a shared silicon platform for fabricating small digital and mixed-signal designs~\citep{venn2024tinytapeout}, while Croc uses an open-source RISC-V platform and design flow to support system-on-chip (SoC) education~\citep{zelioli2026croc}.

Despite these advances, practical exposure alone does not necessarily make the relationships among design representations and workflows easier to understand. EDA environments produce timing reports, layout views, design-rule violations, and other outputs that require substantial technical knowledge to interpret. Understanding how a placement decision affects routing congestion, signal delay, or power, for instance, requires students to connect physical organization with electrical behavior and design constraints. Prior research has highlighted the challenges of teaching complex microelectronics workflows and preparing students for increasingly interconnected design tasks~\citep{ref-journal7,ref-journal9-razavi2024education}. These challenges underscore the need for instructional approaches that support reasoning across representations and help learners interpret the consequences of their decisions.

Research on educator-centered learning technology design offers a complementary perspective. Turakhia et al.~\citep{turakhia2024educators} examined instructional practices in makerspaces to understand how educators support hands-on learning and how educational tools might better address their needs. Although chip design presents different technical challenges, this approach highlights the value of grounding new learning technologies in existing instructional practices rather than introducing technologies independently of the learning problems they are intended to address.

\subsection{Emerging Technologies for Chip Design Learning}

Recent advances in generative artificial intelligence (AI) and large language models (LLMs) have introduced new ways to support hardware design and make complex workflows more accessible. Research has explored translating natural-language specifications into Hardware Description Languages (HDLs), allowing users to express intended circuit behavior without constructing every element of the implementation manually~\citep{ref-journal32-yang2025nls,krupp2026rtlprompt}. Related approaches have investigated LLM assistance for testbench generation, code interpretation, verification, and debugging~\citep{ref-journal35-abdollahi2025hardware}. Such tools may reduce barriers associated with HDL syntax and specialized toolchains, particularly for learners beginning to work with digital hardware.

However, simplifying the production of design artifacts does not necessarily simplify the reasoning required to understand them. A generated RTL description must still be connected to its underlying hardware behavior, synthesized logic, physical implementation, and performance implications. Similarly, identifying or correcting an error in code does not by itself explain how that error affects timing, connectivity, or downstream design constraints. As language-based interfaces increasingly support movement from initial specifications toward more complete hardware implementations~\citep{krupp2026rtlprompt}, understanding the relationships among these representations remains an important educational challenge.

Immersive technologies offer a different, potentially complementary form of support. Rather than primarily assisting with the generation or modification of design artifacts, VR and AR can provide interactive spatial representations through which learners explore structures, manipulate configurations, and examine dynamic processes. Research on immersive learning has investigated how these capabilities relate to conceptual understanding, engagement, and knowledge transfer~\citep{ref-journal36-klingenberg2024embodiment,radianti2020review,ref-journal37-huang2025extended}. For chip design education, their potential lies in making connections between abstract descriptions, physical structures, and design consequences more accessible to learners.

\subsection{Immersive Technologies in STEM Education}

Immersive technologies have been explored extensively in STEM education, particularly where important concepts involve spatial relationships, dynamic processes, or phenomena that are difficult to observe directly. VR and AR can make such phenomena accessible through interactive representations that learners can inspect from different perspectives. For example, immersive visualization has been used to support learning about DNA structures and their relationship to biological function~\citep{ref-proceeding11-dunn2023hands}, while applications in chemistry and physics have examined molecular interactions, heat transfer, and electromagnetic phenomena~\citep{ref-proceeding12-amaral2024interactive,ref-journal13-strzys2018physics,ref-journal14-warkentin2025hands}. These studies illustrate how immersive representations can provide additional ways to engage with concepts that are otherwise communicated through diagrams, equations, or physical demonstrations.

Beyond conceptual learning, immersive environments have also been investigated for procedural training and complex physical skills. In engineering and vocational contexts, VR systems have supported activities such as welding, building inspection, prototyping, and Internet of Things (IoT) installation~\citep{ref-proceeding38-ipsita2022vrwelding,ref-proceeding39-henstrom2023immersive,ref-proceeding40-lee2025immersive,ref-proceeding41-zhu2023learniotvr}. Simulated environments can provide opportunities for repeated practice and experimentation while reducing some of the logistical constraints and risks of working with physical equipment. More broadly, research on three-dimensional virtual environments has identified spatial exploration, interaction, and experiential learning among their potential educational affordances~\citep{dalgarno2010learning,merchant2014effectiveness}.

Work on immersive manufacturing training offers a particularly relevant example of how these capabilities can support learning in technically constrained environments. Lie et al.~\citep{lie2023drilling} developed and evaluated an open-ended VR drilling simulation that allowed learners to practice machining procedures, explore alternative actions, and encounter the consequences of errors. Studer et al.~\citep{studer2024openended} further examined an open-ended VR environment for machining training, emphasizing opportunities for learners to pursue different strategies while working within essential task and safety constraints. Although machining differs from chip design, these efforts demonstrate how immersive systems can be designed around authentic engineering decisions rather than fixed instructional sequences. This perspective is relevant to chip design tasks in which students must explore alternatives, recognize constraints, and understand why particular decisions succeed or fail.

At the same time, immersion does not inherently lead to better learning. The effectiveness of immersive environments depends on how their representations, interactions, and instructional guidance align with the learning task. For example, \citet{makransky2019presence} found that an immersive VR science simulation increased learners' sense of presence but resulted in lower learning performance and higher cognitive load compared with a desktop version. Such findings caution against treating immersion as an educational advantage in itself, particularly when learners must coordinate complex information or master unfamiliar interfaces.

This distinction is especially relevant to chip design education. Understanding semiconductor structures, interpreting timing behavior, navigating design representations, and evaluating implementation trade-offs involve different forms of knowledge and reasoning. While some of these tasks may benefit from three-dimensional exploration or direct manipulation, others may be better supported through conventional diagrams, simulations, or design tools. Identifying the educational value of immersion therefore requires examining the learning demands of specific tasks rather than assuming that the entire design process should be translated into VR.

\subsection{Immersive Technologies for Chip Design Education}

Despite growing research on immersive STEM learning, applications specifically addressing chip design remain relatively limited. Paszkiewicz et al.~\citep{ref-journal16-paszkiewicz2021vr} developed a VR-based educational system for digital circuit design, allowing learners to interact with circuit components and explore introductory logic concepts. Such work illustrates the potential of immersive environments for learning circuit structures and behavior, but addresses only part of the broader chip design process.

As learners progress beyond introductory circuit concepts, they encounter increasingly interconnected design tasks. Architecture, synthesis, floorplanning, placement, routing, timing optimization, and physical verification require students to move between logical and physical representations while accounting for constraints involving power, performance, area, and manufacturability. Fabrication and packaging introduce additional relationships between abstract layouts, physical processes, and device reliability. The nature of these tasks varies considerably: interpreting a multilayer routing structure presents different learning challenges from diagnosing a timing violation or evaluating a floorplanning trade-off. Consequently, the forms of immersive visualization, interaction, and feedback that might support learning are unlikely to be uniform across the pipeline.

Existing research provides valuable precedents for immersive circuit visualization and experiential engineering training, but offers limited guidance on how to select immersive capabilities for the distinct learning demands of chip design. This motivates a shift from developing isolated VR demonstrations toward understanding the educational challenges that such environments should address. Our study takes this learning-centered approach by examining the knowledge and reasoning required across the chip design curriculum, identifying recurring learning difficulties, and mapping these demands to relevant immersive affordances. In doing so, we aim to provide a more systematic foundation for designing immersive learning environments that complement existing chip design instruction.

\section{Methodology}
\label{sec:methodology}

We combined a chip design curriculum analysis with semi-structured interviews to identify learning demands and instructional challenges across the design pipeline. The curriculum analysis characterized the concepts, workflows, and decisions involved in chip design; interviews provided perspectives on how these topics are taught and where learners encounter difficulty. We then synthesized both sources and developed a conceptual mapping between identified learning demands and potential immersive affordances.

\subsection{Curriculum Analysis}

We examined chip design curricula spanning semiconductor fundamentals, digital logic and architecture, register-transfer-level (RTL) design, functional verification, logic synthesis, physical implementation, and manufacturing. We organized these topics into three curricular areas: \textit{device physics and circuit foundations}, \textit{the chip design cycle}, and \textit{tapeout and manufacturing}. Within each area, we characterized the concepts learners must understand, the workflows they must perform, and the relationships and trade-offs they must evaluate. Particular attention was given to transitions between representations---such as RTL, netlists, timing reports, and layouts---and to the downstream consequences of design decisions. This analysis established the learning units used to organize the interview findings and subsequent affordance mapping.


\subsection{Participant Recruitment and Backgrounds}

We used purposive sampling, initially contacting faculty at a research-oriented engineering institution and expanding recruitment through referrals. Five participants contributed perspectives on chip design instruction and learning: three faculty members teaching VLSI, system-on-chip (SoC), or digital logic, and two advanced learners with prior teaching-assistant experience. Participants were affiliated with or closely connected to the same engineering education context. Table~\ref{tbl:interviewees} summarizes their backgrounds. The sample was intended to support an exploratory account of this curricular context, not statistical representation of chip design programs.

\begin{table*}[h!]
\caption{Participant backgrounds and perspectives on chip design education.}
\label{tbl:interviewees}
\small
\setlength{\tabcolsep}{4pt}
\renewcommand{\arraystretch}{1.15}
\begin{tabular*}{\textwidth}{@{\extracolsep{\fill}}
>{\raggedright\arraybackslash}p{0.06\textwidth}
>{\raggedright\arraybackslash}p{0.18\textwidth}
>{\raggedright\arraybackslash}p{0.28\textwidth}
>{\raggedright\arraybackslash}p{0.36\textwidth}@{}}
\toprule
\textbf{ID} & \textbf{Role} & \textbf{Relevant experience} & \textbf{Primary perspective} \\
\midrule
E1 & Professor & Graduate VLSI teaching and curriculum development & Abstraction levels and physical-design trade-offs \\
E2 & Assistant Professor & Advanced SoC design and EDA workflows & Architecture-to-implementation relationships \\
E3 & PhD Candidate / Former TA & VLSI instruction and chip design research & Difficulties interpreting representations and tool outputs \\
E4 & Graduate Student / Former TA & Advanced VLSI coursework and embedded-systems teaching & Recent learner experience with chip design workflows \\
E5 & Professor & Digital logic, hardware design, and AI-supported instruction & Curriculum design and emerging instructional tools \\
\bottomrule
\end{tabular*}
\end{table*}

\subsection{Semi-Structured Interviews}

We conducted five semi-structured interviews, each lasting approximately 45--60 minutes. Interviews were audio-recorded and transcribed. The protocol covered three areas: (1) participants' instructional or learning experiences, including teaching methods, tools, and assessment practices; (2) difficult concepts and transitions across the chip design pipeline, including spatial layouts, timing behavior, and design consequences; and (3) possible uses of VR and AR, such as layout visualization, fabrication walkthroughs, debugging, and design feedback. Follow-up questions elicited concrete examples from participants' experiences. The interviews explored perceived needs and opportunities; participants did not evaluate an immersive prototype.


\subsection{Data Analysis and Synthesis}

Our analysis proceeded in two stages: thematic synthesis of learning demands and challenges, followed by conceptual mapping to immersive affordances.

\subsubsection{Thematic coding and synthesis}

We used a hybrid inductive--deductive thematic approach~\citep{ref-journal19-naeem2023step}. Through iterative transcript review and open coding, we identified accounts of instructional practices, difficult representations, and reasoning across design stages. Related codes were grouped into recurring challenges and considered alongside the curriculum analysis. Informed by learning taxonomies and design-oriented scholarship~\citep{ref-book20-anderson2001taxonomy,ref-book21-cross2006designerly}, we organized learning demands into four complementary categories: \textit{descriptive knowledge} (concepts and structures), \textit{procedural knowledge} (design workflows), \textit{relational and design-focused reasoning} (connections across representations, components, and decisions), and \textit{strategic decision-making} (alternatives and trade-offs under constraints). The categories provided a common lens for interpreting demands across curricular units; spatial, temporal, and diagnostic reasoning were treated as cross-cutting skills rather than additional categories.


\subsubsection{Mapping learning demands to immersive affordances}

We then examined which immersive capabilities could plausibly support the reasoning required by each curriculum unit. Drawing on immersive-learning research~\citep{dalgarno2010learning,radianti2020review,ref-journal23-parong2018learning,ref-journal24-bailenson2008use}, we considered six affordances: spatial visualization, multi-scale navigation, interactive manipulation, temporal dynamics visualization, immediate feedback, and collaboration. We linked learning challenges to affordances based on the representations and interactions they might require. For example, interpreting multilayer interconnects motivated spatial visualization and cross-layer navigation; examining signal propagation motivated temporal visualization; and evaluating placement or routing alternatives motivated manipulation with feedback. These associations constitute a researcher-developed conceptual mapping informed by the curriculum, interviews, and literature---not participant-endorsed prescriptions or evidence that VR improves learning.


\section{Results and Analysis}

The analysis begins with the curriculum and interviews, which reveal not only the breadth of chip design knowledge but also the difficulty of connecting that knowledge across representations and stages. We then consider the immersive capabilities relevant to these difficulties and develop a task-level mapping between learning demands and potential design responses. The mapping is an analytical proposal, not evidence that immersive instruction improves learning.


\definecolor{ChipHeader}{HTML}{DFEAF3}
\definecolor{ChipStripe}{HTML}{F3F7FA}

\subsection{Learning Demands and Challenges in Chip Design Education}
\label{sec:learning-demands}

Across the curriculum, the difficulty of chip design lies as much in the connections between topics as in the topics themselves. Students encounter semiconductor physics, logic design, physical implementation, and manufacturing through different representations and tools; yet effective design reasoning depends on understanding how those views relate. We use the curriculum analysis to identify the demands at each stage and the interviews to examine where these connections become difficult for learners.

\subsubsection{Curriculum Structure and Learning Scope}
\label{sec:curriculum-structure}

The curriculum was organized into three connected areas: \textit{device physics and circuit foundations}, \textit{the chip design cycle}, and \textit{tapeout and manufacturing}. In the first, learners connect material properties and transistor geometry to electrical behavior and logic functions. The design cycle moves between specifications, RTL, netlists, timing reports, and physical layouts, demanding increasingly close attention to how one representation constrains another. Manufacturing extends these relationships to process sequences, testing, packaging, and reliability. Tables~\ref{tab:curriculum-device}--\ref{tab:curriculum-manufacturing} trace the core concepts and reasoning challenges within these areas. The demand labels D, P, R, and S refer to descriptive, procedural, relational and design-focused, and strategic learning demands; each row identifies prominent rather than exclusive demands. Full curriculum inventories are retained as supplementary material.

Although presented in stages, design is not a one-way progression. Fabrication rules shape possible layouts; placement and routing alter timing outcomes; and an optimization made late in the flow may require revisiting earlier decisions. The curriculum therefore asks students to understand both the operation of individual tools and the dependencies that link them.

\begin{table*}[t]
\centering
\caption{Device physics and circuit foundations: key concepts, dominant learning demands, and reasoning challenges.}
\label{tab:curriculum-device}
\small
\setlength{\tabcolsep}{4pt}
\renewcommand{\arraystretch}{1.17}
\begin{tabularx}{\textwidth}{@{}>{\raggedright\arraybackslash}p{0.17\textwidth}>{\raggedright\arraybackslash}X>{\raggedright\arraybackslash}p{0.12\textwidth}>{\raggedright\arraybackslash}X@{}}
\toprule
\rowcolor{ChipHeader}
\textbf{Learning unit} & \textbf{Key concepts and workflows} & \textbf{Dominant demands} & \textbf{Central reasoning challenge} \\
\midrule

\rowcolor{ChipStripe} Semiconductor physics & Doping, charge transport, electric fields, device equations & D, R & Connect material properties and carrier behavior to device characteristics.\\
Transistor level & MOSFET/FinFET operation, device geometry, I--V behavior, SPICE and sizing & D, R, S & Relate structure and parameters to switching speed, leakage, and power.\\
\rowcolor{ChipStripe} Standard cells & Transistor-level logic, cell libraries, drive-strength variants, layout constraints & R, S & Explain how transistor organization and cell selection affect logic, timing, and layout.\\
\bottomrule
\end{tabularx}
\end{table*}

\begin{table*}[!t]
\centering
\caption{Chip design cycle: key concepts, dominant learning demands, and reasoning challenges across specification, front-end design, physical implementation, and verification.}
\label{tab:curriculum-frontend}
\small
\setlength{\tabcolsep}{4pt}
\renewcommand{\arraystretch}{1.17}
\begin{tabularx}{\textwidth}{@{}>{\raggedright\arraybackslash}p{0.17\textwidth}>{\raggedright\arraybackslash}X>{\raggedright\arraybackslash}p{0.12\textwidth}>{\raggedright\arraybackslash}X@{}}
\toprule
\rowcolor{ChipHeader}
\textbf{Learning unit} & \textbf{Key concepts and workflows} & \textbf{Dominant demands} & \textbf{Central reasoning challenge} \\
\midrule

\rowcolor{ChipStripe} Specification and architecture & Functional requirements, block organization, dataflow, PPA targets & R, S & Translate system requirements into architectural and implementation choices.\\
RTL design & HDL, registers, datapaths, control, simulation & P, R & Express specifications as concurrent hardware behavior and trace resulting signals.\\
\rowcolor{ChipStripe} Functional verification & Testbenches, assertions, simulation, coverage, debugging & P, S & Connect test outcomes and coverage gaps to functional correctness.\\
Logic synthesis & RTL-to-gate mapping, constraints, optimization, timing reports & P, R, S & Link coding and synthesis choices to netlist structure and PPA outcomes.\\
\rowcolor{ChipStripe} Floorplanning & Macro placement, I/O planning, power grid, blockages & R, S & Anticipate how block arrangement affects congestion, timing, and power.\\
Cell placement & Standard-cell positioning, legalization, wirelength and congestion optimization & P, R, S & Relate local cell positions to routing demand and path delay.\\
\rowcolor{ChipStripe} Clock-tree synthesis & Buffering, clock distribution, skew and insertion delay & R, S & Connect clock-network geometry to arrival times, skew, and power.\\
Timing optimization & Setup/hold analysis, critical paths, sizing, buffering, logic changes & P, R, S & Diagnose violations and evaluate fixes that affect multiple constraints.\\
\rowcolor{ChipStripe} Routing & Global/detailed routing, vias, metal layers, parasitics and DRC & P, R, S & Trace multilayer connectivity and explain congestion, delay, or spacing problems.\\
Finishing and signoff preparation & Placement cleanup, engineering changes, final checks and fixes & P, S & Prioritize late-stage corrections without creating new physical violations.\\
\rowcolor{ChipStripe} Physical verification & DRC, LVS, ERC, connectivity, extraction and manufacturability checks & P, R, S & Map violations to geometry or connectivity and identify corrective actions.\\
\bottomrule
\end{tabularx}
\end{table*}

\begin{table*}[!t]
\centering
\caption{Tapeout and manufacturing: key concepts, dominant learning demands, and reasoning challenges.}
\label{tab:curriculum-manufacturing}
\small
\setlength{\tabcolsep}{4pt}
\renewcommand{\arraystretch}{1.17}
\begin{tabularx}{\textwidth}{@{}>{\raggedright\arraybackslash}p{0.17\textwidth}>{\raggedright\arraybackslash}X>{\raggedright\arraybackslash}p{0.12\textwidth}>{\raggedright\arraybackslash}X@{}}
\toprule
\rowcolor{ChipHeader}
\textbf{Learning unit} & \textbf{Key concepts and workflows} & \textbf{Dominant demands} & \textbf{Central reasoning challenge} \\
\midrule

\rowcolor{ChipStripe} Fabrication & Lithography, deposition, etching, doping, process variation & D, P, R & Link layouts and process sequences to formed structures and yield.\\
Testing and validation & Test vectors, DFT, fault detection, burn-in, reliability testing & P, R, S & Interpret failures and balance test coverage with time and cost.\\
\rowcolor{ChipStripe} Packaging and deployment & Die attachment, wire bonding/flip-chip, thermal and I/O integration & R, S & Relate package architecture to signal integrity, thermal behavior, and reliability.\\
\bottomrule
\end{tabularx}
\end{table*}

\subsubsection{Learning Demands Across the Design Pipeline}
\label{sec:learning-goals}

The curriculum exposes four intertwined forms of learning. \textit{Descriptive knowledge} provides the vocabulary and models needed to explain devices, circuits, and design structures. \textit{Procedural knowledge} concerns carrying out design steps and interpreting what tools produce. \textit{Relational and design-focused reasoning} connects representations, components, and stages, while \textit{strategic decision-making} involves choosing among alternatives under constraints on power, performance, area, reliability, and manufacturability. These categories describe what a task demands rather than separate kinds of students or isolated curricular units.

Their overlap becomes clearest when a learner moves from a formal specification to an implemented design. E2 described how even the initial translation can be demanding:

\begin{quote}
\small
\textit{``Given an English specification, being able to formulate the problem as a Boolean algebra problem, and then as a digital logic problem, both as combinational logic and as sequential logic, becomes complicated.''} \hfill (E2)
\end{quote}

The difficulty does not disappear once a representation has been produced. RTL must be checked for function, synthesized into gates, and interpreted against timing and physical constraints. Device physics places greater weight on concepts and structural relationships; verification and implementation add procedural and strategic judgments; manufacturing brings process and reliability considerations into view. Spatial, temporal, and diagnostic reasoning run through these activities rather than forming additional, competing categories. For example, diagnosing a critical path requires interpreting a timing report, locating the corresponding logic and interconnects, and deciding which correction is appropriate.

The interviews help explain why these intertwined demands are difficult in practice. We distinguish five recurrent challenges, recognizing that participant accounts describe experiences and interpretations rather than direct measures of learning performance.

\subsubsection{Connecting Abstraction Levels}
\label{sec:learning-challenges}
A chip is encountered through descriptions that reveal different parts of the same underlying design: natural-language requirements, Boolean expressions, RTL, gate-level netlists, physical layouts, and reports. Learners must understand what each representation preserves, what it abstracts away, and what new constraints appear as the design becomes more concrete. E1 contrasted the relative simplicity of introductory models with the non-idealities and implementation decisions that emerge in practical workflows.
For example, a student may correctly explain an RTL module's function without being able to anticipate how synthesis will map it to gates or why two functionally equivalent implementations differ in timing and area. What makes the transition challenging is not simply unfamiliar syntax or a new tool. It is the need to recognize continuity of function amid changes in structure, constraints, and scale. This helps explain why mastering one stage of the curriculum does not necessarily translate into fluency in the next.

\subsubsection{Interpreting Spatial Structures}
Spatial interpretation introduces another kind of translation. Schematics abstract away geometry, whereas layout views encode devices and connectivity as shapes distributed across layers. Students must infer how transistor regions, cell boundaries, metal lines, and vias collectively implement the circuits represented symbolically. As E3 explained:

\begin{quote}
\small
\textit{``Trying to understand and make the switch from logical to physical is always a challenge. How do you take this design that is on paper and translate it into geometric shapes on a piece of silicon that will be manufactured?''} \hfill (E3)
\end{quote}

E2 similarly discussed the difficulty of relating gate symbols and stick diagrams to transistor structures and their interconnections. Multilayer routing makes the problem concrete: a connection that is continuous electrically may be fragmented across layer-specific drawings, leaving learners to reconstruct its path through vias and metal segments. Thus, the underlying difficulty is not merely the inability to imagine a three-dimensional object; it is coordinating geometry, connectivity, and function as different descriptions of that object.

\subsubsection{Understanding Temporal Behavior}
Even when circuit structures are understood, their behavior depends on events that occur over time. Signals propagate along paths with different delays, sequential elements depend on clock relationships, and fabrication processes build structures through ordered operations. Timing diagrams and simulation reports make these behaviors available for inspection, but understanding them requires connecting abstract events to the circuit that produces them. E2 contrasted the habitual expectations of software programming with the concurrency of hardware:

\begin{quote}
\small
\textit{``When you do programming, everything is sequential. You write a program, and each instruction executes, followed by the next instruction. But with gates, these are independent events.''} \hfill (E2)
\end{quote}

Although software may also execute concurrently, the comparison points to a specific teaching difficulty: the learner must track events that overlap and interact rather than follow a single sequence of instructions. A setup violation, for example, cannot be diagnosed by reading arrival times alone; it must be related to the data path, clock path, and relevant constraints. Temporal understanding is therefore inseparable from structural and representational understanding.

\subsubsection{Connecting Theory with Practice}
Participants also distinguished knowing a principle from recognizing when and how to use it. E1 emphasized how practical flows introduce effects not visible in simplified theoretical models, while E3 described how concepts taught in lectures often acquire meaning once students encounter a corresponding design problem. E4 noted that students arrive at advanced VLSI coursework with uneven hardware preparation, which affects how readily they can engage with such tasks.
A learner might know the definition of a critical path yet struggle to locate it in a timing report, connect it to the physical layout, and determine whether buffering, sizing, or placement is an appropriate response. Authentic design work demands coordination of concepts, tools, and judgment, often under conditions that differ from classroom examples. The accounts point to a gap in the application and integration of knowledge, although they do not establish that any particular instructional intervention closes it.

\subsubsection{Reasoning about Trade-offs and Downstream Consequences}
Finally, chip design rarely permits a change to be evaluated against a single objective. Increasing cell drive strength may improve delay but affect power and area; rerouting may resolve congestion locally while altering parasitics elsewhere; and decisions made during floorplanning can constrain subsequent clock distribution and routing. The relevant expertise lies in recognizing these dependencies and justifying a choice in light of competing constraints.
This theme is clearest in the curriculum analysis of physical implementation and verification. The interviews offer broader accounts of the theory--practice divide but less direct evidence about how individual learners weigh design trade-offs. We therefore treat downstream consequence reasoning primarily as a curriculum-derived demand rather than as a separately demonstrated learner deficit. Its importance is nonetheless evident: students must learn to explain not only whether a design change works locally, but what it may cost elsewhere.

\subsubsection{Synthesis: Reasoning Across Scales}
\label{sec:cross-scale-synthesis}
Taken together, the findings point toward a recurring demand to move between physical structures, symbolic representations, temporal behavior, and decisions whose effects cross design stages. A routing problem, for example, may require reading a layout geometrically, tracing connectivity electrically, interpreting design rules, and judging whether a correction introduces new timing consequences. The challenge is not a single missing skill but the coordination of several forms of reasoning.
This interpretation provides a more useful starting point for immersive learning than a general claim that chip design is spatial or visually complex. It directs attention to the specific relationships learners need to reconstruct and asks which forms of representation and interaction might help them do so.

\subsection{Immersive Affordances for Chip Design Learning}
\label{sec:immersive-affordances}

The learning challenges described above involve relationships that are often distributed across diagrams, tool outputs, and stages of instruction. Immersive technologies offer ways of bringing some of those relationships into a shared representational space, though their educational value depends on the task. Informed by research on virtual learning environments~\citep{dalgarno2010learning,radianti2020review,ref-journal23-parong2018learning,ref-journal24-bailenson2008use}, we consider six affordances: spatial visualization, multi-scale navigation, interactive manipulation, temporal dynamics visualization, immediate feedback, and collaboration. Table~\ref{tbl:affordance_taxonomy} distinguishes these capabilities; the discussion below considers what they might contribute to chip design learning and where their limitations matter.

\begin{table*}[h!]
\centering
\caption{Immersive affordances and their potential applications in chip design education.}
\label{tbl:affordance_taxonomy}
\small
\renewcommand{\arraystretch}{1.18}
\setlength{\tabcolsep}{5pt}

\begin{tabularx}{\textwidth}{@{}
>{\raggedright\arraybackslash}p{0.19\textwidth}
>{\raggedright\arraybackslash}X
>{\raggedright\arraybackslash}p{0.32\textwidth}
@{}}
\toprule
\rowcolor{blue!12}
\textbf{Affordance} &
\textbf{Description} &
\textbf{Example in Chip Design} \\
\midrule

Spatial visualization &
Making complex three-dimensional structures and spatial relationships visible and explorable. &
Inspecting transistor geometries, standard cells, and multilayer interconnects. \\
\rowcolor{blue!4}

Multi-scale navigation &
Moving between physical scales, abstraction levels, and related design representations. &
Exploring connections between transistor structures, logic cells, functional blocks, and chip layouts. \\

Interactive manipulation &
Modifying design representations or parameters to explore alternative configurations and their effects. &
Adjusting floorplans, cell placement, routing paths, or device parameters. \\
\rowcolor{blue!4}

Temporal dynamics visualization &
Making changes over time visible, including signal propagation, synchronization, and process sequences. &
Visualizing clock propagation, timing violations, or fabrication steps. \\

Immediate feedback &
Providing interpretable responses to learner actions to reveal relationships between choices and outcomes. &
Showing changes in timing, congestion, power, or design-rule violations following design modifications. \\
\rowcolor{blue!4}

Collaboration &
Supporting shared representations and coordinated exploration of design problems. &
Conducting collaborative layout reviews, debugging, and design trade-off discussions. \\

\bottomrule
\end{tabularx}
\end{table*}

\subsubsection{Spatial Visualization}
Spatial visualization can make device geometries and physical relationships accessible when they are distributed across diagrams, cross-sections, and layout views. For example, students could examine the arrangement of transistor regions, the organization of devices within standard cells, or the connections between wires and vias across metal layers. The educational value is not simply that the objects appear in three dimensions, but that their relative positions, continuity, and functional relationships become easier to inspect.
Such representations would need to be tied to engineering meaning. A visually detailed transistor model is of limited value unless learners can relate its geometry to device operation; similarly, an elaborate layout view must still communicate connectivity and constraints. Selective highlighting, cutaway views, and links to schematic representations may therefore be more useful than visual realism alone.

\subsubsection{Multi-Scale Navigation}
Learning across the chip design pipeline requires movement between transistors, standard cells, blocks, and whole-chip layouts, as well as between different representations of the same design. Immersive navigation could help preserve correspondences across these transitions: a learner might select a cell in a layout and inspect its logical function or trace how a net appears in both a schematic and its physical routing. This is more demanding than zooming, because a change in abstraction often changes what information is represented.

E2 described the difficulty of following paths across physical layers:
\begin{quote}
\small
\textit{``To connect gates, you have to go from one layer to the next layer, walk through that second layer, and then somehow come back to the previous layer.''} \hfill (E2)
\end{quote}

This observation suggests value in navigation that retains orientation and connectivity while views or layers change. However, learners must also be able to return to the two-dimensional and symbolic forms used in professional design tools; immersion should help bridge those representations rather than isolate them.

\subsubsection{Interactive Manipulation}
Interactive manipulation could allow students to modify device parameters, reorganize a floorplan, reposition cells, or explore alternative routing paths. These actions create opportunities to test predictions and compare configurations, particularly where students otherwise see only a completed result or a sequence of tool commands. For example, moving a macro could provide a concrete starting point for considering how wire length, congestion, and timing depend on placement.
The interaction must nevertheless be constrained by the engineering problem. Arbitrary spatial changes without valid design rules or interpretable outcomes risk teaching misleading causal relationships. Appropriate activities might therefore use simplified design spaces, guided alternatives, or links to analysis outputs, depending on the learner's preparation and the instructional objective.

\subsubsection{Temporal Dynamics Visualization}
Temporal dynamics visualization can connect changing behavior to the structures that produce it. In chip design, this includes signal propagation through combinational logic, clock distribution across sequential elements, setup and hold relationships, and the ordered operations of fabrication. Conventional timing diagrams and process illustrations express these phenomena, but learners must mentally align them with the corresponding circuit or material structures.
An immersive timing activity could show signal arrival along a clock network while retaining a synchronized timing diagram, enabling students to relate path delay to skew and synchronization. Likewise, a fabrication walkthrough could reveal how successive deposition, patterning, and etching steps produce layered structures. These animations would need accurate timing or process models, with clear distinctions between physically meaningful behavior and schematic visual explanation.

\subsubsection{Immediate Feedback}
Immediate feedback could reveal how a learner's modification affects performance or compliance. Changes to placement, routing, or sizing might be accompanied by interpretable updates to estimated timing, congestion, power, or design-rule violations. E3 expressed interest in tools that enable simple experimentation:
\begin{quote}
\small
\textit{``If you can play around with this simple tool, where you move some sliders around and it shows you what the effect is, then that could be helpful.''} \hfill (E3)
\end{quote}

The pedagogical opportunity is to support comparison between an anticipated and an observed outcome, rather than merely signal that an action was correct or incorrect. Feedback must also expose relevant assumptions: a simplified model can show directional trade-offs, whereas accurate analyses may require external simulators or EDA tools and may not update instantaneously. Making this distinction visible is important for avoiding false confidence in the represented results.

\subsubsection{Collaboration}
Shared immersive representations could support design reviews in which students and instructors inspect a layout together, identify problematic regions, trace paths, and discuss alternative corrections. A common spatial reference may be particularly helpful when participants are trying to explain a complex multilayer feature or compare several physical-design strategies. Annotations and synchronized views could also help connect verbal explanations to specific elements of the design.
Collaboration is not unique to immersive environments, and desktop design reviews already support many of these activities. The relevant opportunity is therefore whether shared spatial interaction offers additional clarity or coordination for particular tasks. Learning activities should preserve the ability to explain decisions with familiar engineering artifacts and should be evaluated against suitable non-immersive alternatives.
These affordances are complementary rather than interchangeable. Navigation is most useful when it preserves relationships across views; manipulation becomes educationally meaningful when consequences can be interpreted; and feedback requires models that faithfully represent the constraints being taught. None of these capabilities is exclusive to immersive interfaces. The relevant question is whether a spatially integrated interaction offers an advantage for a particular learning task over existing diagrams, simulations, and EDA tools.

\subsection{Mapping Learning Demands to Immersive Affordances}
\label{sec:affordance-mapping}

The preceding analysis suggests that an immersive learning activity should be selected because of the reasoning it supports, not simply because a topic can be rendered in three dimensions. We therefore mapped representative curriculum tasks to the affordances most relevant to their dominant learning demands and observed or curriculum-derived challenges. Table~\ref{tbl:learning-affordance-mapping} shows this researcher-developed synthesis, identifying possible supports rather than empirically established learning effects. The mapping is selective: the listed demands and affordances are particularly salient for each task, not exhaustive.

\begin{table*}[!t]
\centering
\caption{Mapping learning demands and reasoning challenges across the chip design pipeline to relevant immersive affordances. The mapping identifies proposed educational opportunities rather than empirically validated interventions.}
\label{tbl:learning-affordance-mapping}
\small
\renewcommand{\arraystretch}{1.18}
\setlength{\tabcolsep}{4pt}

\begin{tabularx}{\textwidth}{@{}
>{\raggedright\arraybackslash}p{0.17\textwidth}
>{\raggedright\arraybackslash}p{0.19\textwidth}
>{\raggedright\arraybackslash}X
>{\raggedright\arraybackslash}p{0.24\textwidth}
@{}}
\toprule
\rowcolor{blue!12}
\textbf{Learning task} &
\textbf{Dominant learning demands} &
\textbf{Key reasoning challenge} &
\textbf{Relevant immersive affordances} \\
\midrule

Semiconductor and transistor behavior &
Descriptive; relational &
Connecting device geometry, material properties, and electrical behavior. &
Spatial visualization; interactive manipulation \\

\rowcolor{blue!4}
Standard-cell structures &
Descriptive; relational &
Understanding how transistor arrangements form logic cells and influence circuit behavior. &
Spatial visualization; multi-scale navigation \\

Specification and RTL design &
Procedural; relational &
Translating functional specifications into hardware structures and behavior. &
Multi-scale navigation; temporal dynamics visualization \\

\rowcolor{blue!4}
Logic synthesis &
Procedural; relational &
Connecting RTL descriptions to gate-level structures and implementation constraints. &
Multi-scale navigation; spatial visualization \\

Floorplanning and placement &
Relational; strategic &
Understanding how block organization and cell placement affect timing, congestion, and power. &
Spatial visualization; interactive manipulation; immediate feedback \\

\rowcolor{blue!4}
Clock-tree synthesis &
Relational; strategic &
Understanding clock propagation, synchronization, skew, and insertion delay. &
Temporal dynamics visualization; immediate feedback \\

Timing optimization &
Relational; strategic &
Tracing critical paths and evaluating how timing fixes affect other design constraints. &
Temporal dynamics visualization; interactive manipulation; immediate feedback \\

\rowcolor{blue!4}
Multilayer routing &
Procedural; relational &
Understanding connectivity across metal layers, vias, and routing constraints. &
Spatial visualization; multi-scale navigation; interactive manipulation \\

Physical verification &
Procedural; strategic &
Identifying design-rule violations, tracing their causes, and evaluating corrective actions. &
Spatial visualization; interactive manipulation; immediate feedback \\

\rowcolor{blue!4}
Fabrication and manufacturing &
Descriptive; procedural &
Connecting layout representations to fabrication sequences and resulting physical structures. &
Spatial visualization; temporal dynamics visualization; multi-scale navigation \\

Testing and packaging &
Relational; strategic &
Connecting physical implementation, testing, and packaging decisions to reliability and system behavior. &
Spatial visualization; multi-scale navigation; collaboration \\

\bottomrule
\end{tabularx}
\end{table*}

The mapping is most revealing where a task requires several kinds of reasoning at once. Consider timing optimization: a learner must connect a logical path to its physical implementation, understand how delay produces a constraint violation, and assess how a proposed fix affects power, area, or routing. A spatial view alone would not address this task. Its instructional promise lies in coupling path visualization to temporal behavior and interpretable feedback about alternatives.

A similar pattern emerges when students move between design representations. RTL, synthesized logic, and routed layouts do not merely present the same information at different zoom levels; each introduces different structures and constraints. An environment that preserves correspondences between these representations may help learners identify what changes during implementation and what remains functionally equivalent. At the device level, by contrast, the immediate need may be to relate geometry to electrical operation, making spatial inspection and controlled parameter exploration more relevant.

These examples also mark the limits of the mapping. Collaboration, direct manipulation, and immediate feedback are possible in desktop tools, while immersive interaction may introduce its own demands. The framework therefore identifies candidate opportunities to investigate rather than a prescribed VR curriculum. Its practical value lies in specifying which relationships an activity should make accessible and which reasoning outcomes a future evaluation would need to test.

\section{Discussion}
\label{sec:discussion}

The findings point to a gap between learning the individual components of chip design and understanding how those components fit together. Students encounter semiconductor devices, logic design, timing, physical implementation, and fabrication through different representations and tools, yet many design problems require them to reason across these boundaries. The affordance mapping developed in Section~\ref{sec:affordance-mapping} suggests ways immersive learning environments could make such relationships easier to investigate. We discuss what these findings imply for chip design learning, outline possible learning experiences, and consider the challenges involved in developing and evaluating them.

\subsection{From Isolated Concepts to Cross-Scale Reasoning}
\label{sec:discussion-reasoning}

Understanding chip design involves more than progressing through a sequence of concepts and tools. A learner may understand Boolean logic, transistor operation, and timing analysis separately but still struggle to explain how an RTL description is transformed into a physical implementation, or why a change in layout affects circuit performance. These difficulties arise because each representation reveals some properties of a design while obscuring others. Moving between them requires learners to recognize what remains consistent, what changes, and which new constraints become relevant.

We use \textit{Cross-Scale Spatial Reasoning} (CSSR) as a conceptual lens for this interconnected form of reasoning: the ability to relate physical structures, design representations, and their functional consequences across scales and stages of chip design. The term emphasizes spatial relationships but does not reduce the problem to three-dimensional visualization. Interpreting a clock distribution network, for instance, involves relating its physical topology to signal arrival times and skew. Evaluating a floorplan requires connecting block placement to routing demand, timing, and power. In both cases, understanding emerges from coordinating spatial information with behavior and constraints.

This interpretation builds on research showing that learning with multiple representations depends on understanding the relationships among them~\citep{ainsworth2006deft}. Chip design adds a distinctive engineering dimension: representations are connected not only conceptually but through an implementation process in which one decision can introduce constraints elsewhere. Improving delay by resizing a cell, for example, may affect power consumption and placement; rerouting an interconnect may change parasitic effects and congestion. Such dependencies help explain why knowledge acquired through simplified examples does not always transfer readily to realistic design problems.

The educational implication is to teach these connections explicitly. Rather than presenting a detailed transistor model, a timing report, or a layout as separate learning objects, an activity could ask students to predict how a change in one view will appear in another, trace an observed problem to a design decision, or compare solutions against competing constraints. CSSR is proposed here to organize these demands, not as a validated cognitive construct. Its dimensions and relationship to established measures of spatial and representational reasoning remain open to investigation.

\subsection{Design Opportunities for Immersive Chip Design Learning}
\label{sec:discussion-opportunities}

The central opportunity is not to move the entire EDA workflow into VR, but to use immersive interaction selectively where conventional representations make relationships difficult to follow. Table~\ref{tbl:design-opportunities} outlines possible activities and the understanding each is intended to develop. These are design directions derived from our analysis, not implemented or evaluated systems.

\begin{table*}[!t]
\centering
\caption{Illustrative design opportunities for immersive chip design learning. The proposed activities translate identified learning challenges into concrete interactions and intended learning outcomes.}
\label{tbl:design-opportunities}

\small
\renewcommand{\arraystretch}{1.18}
\setlength{\tabcolsep}{4pt}

\begin{tabularx}{\textwidth}{@{}
>{\raggedright\arraybackslash}p{0.18\textwidth}
>{\raggedright\arraybackslash}X
>{\raggedright\arraybackslash}p{0.30\textwidth}
@{}}
\toprule
\rowcolor{blue!12}
\textbf{Learning context} &
\textbf{Proposed immersive interaction} &
\textbf{Intended learning outcome} \\
\midrule

Device physics and transistor structures &
Explore cross-sectional device geometries, inspect semiconductor layers, and adjust parameters while examining simulated electrical characteristics. &
Connect device structure and operating conditions to electrical behavior. \\

\rowcolor{blue!4}
Standard cells and logic synthesis &
Navigate between transistor arrangements, standard-cell layouts, gate-level netlists, and corresponding logic functions. &
Explain how functional descriptions relate to physical circuit implementations. \\

Floorplanning and placement &
Reposition blocks or cells within a spatial layout while inspecting changes in congestion, estimated wire length, timing, and power-related constraints. &
Compare design alternatives and justify trade-offs between competing objectives. \\

\rowcolor{blue!4}
Clock-tree synthesis and timing &
Trace clock and data paths through a layout, visualize arrival times and skew, and investigate how buffering or placement changes affect timing constraints. &
Relate physical implementation choices to temporal behavior and timing violations. \\

Multilayer routing &
Explore metal layers and vias in three dimensions, trace connectivity, and modify routing paths while inspecting spacing constraints and congestion. &
Interpret multilayer connectivity and diagnose routing-related problems. \\

\rowcolor{blue!4}
Physical verification &
Inspect localized design-rule violations and connectivity mismatches, test possible corrections, and compare the resulting verification outcomes. &
Identify error causes and evaluate corrective actions in relation to design constraints. \\

Fabrication and manufacturing &
Step through lithography, deposition, etching, and doping operations while observing how successive processes produce physical structures. &
Connect layout features and manufacturing sequences to fabricated devices and design rules. \\

\rowcolor{blue!4}
Collaborative design review &
Jointly inspect a design, annotate problem areas, compare alternative solutions, and explain design choices within a shared spatial representation. &
Develop evidence-based design explanations and collaborative problem-solving skills. \\

\bottomrule
\end{tabularx}
\end{table*}

\subsubsection{Connecting Physical Structures and Design Representations}

Students often encounter device cross-sections, circuit diagrams, netlists, and layouts independently, even though these describe related aspects of the same design. An immersive activity could let learners examine a transistor structure, identify its role within a standard cell, and then see where that cell appears in a larger circuit. At the design-flow level, selecting an RTL module might reveal the corresponding gate-level structure and physical placement. Such transitions would be useful only if the system makes the correspondence between representations intelligible; enlarging or zooming into a model is not, by itself, an explanation.

These environments could turn the connections into problems for learners to solve. Students might predict which components implement a logic function, follow a signal from its source representation to its physical path, or explain why the same design looks different in a schematic and layout. Coordinated highlighting can guide initial exploration, but activities should ultimately return students to the representations used in actual design practice.

\subsubsection{Making Constraints and Dynamic Behavior Visible}

Many consequences of chip design decisions are difficult to infer from geometry. Timing slack, congestion, parasitic delay, power density, and design-rule violations are reported through specialized views that require experience to interpret. Linking those reports to a spatial representation could help learners locate the relevant structures and reason about their behavior. For example, students might follow a critical timing path through the physical layout while comparing path delay with setup constraints, or inspect how congestion changes around a group of placed cells.

Temporal representations would be especially useful for clock distribution and manufacturing. Showing relative arrival times at clock sinks could clarify the relationship between tree topology and skew, while stepping through lithography and etching could connect processing operations to the formation of physical layers. Such displays must distinguish explanatory animations from simulated or measured phenomena. An intuitive visual encoding is valuable only if students can connect it to the corresponding engineering quantities.

\subsubsection{Supporting Design Exploration and Adaptive Guidance}

Design tasks offer opportunities to learn by making and revising decisions. In a floorplanning exercise, students could move blocks, predict the effects on routing demand, and compare alternatives using estimated timing and congestion. In a timing exercise, they could investigate the effects of cell sizing or buffer insertion while considering the resulting power and area costs. The objective would be to develop explanations for why a design choice works, not merely to improve a score displayed by the interface.

Research on open-ended VR machining has explored how learners can choose among alternative strategies while respecting essential task constraints~\citep{studer2024openended,patel2024benchmarking}. This suggests a useful principle for chip design education: allow meaningful choices, but make functional requirements and physical constraints explicit. Depending on the task, feedback may come from simplified models, precomputed scenarios, or EDA analysis. Its source and accuracy should be transparent to learners.

Guidance could also change with demonstrated progress. Prior work has investigated cognitive architectures for VR tutoring and performance-adaptive AR training~\citep{sepulveda2024actr,turakhia2025adaptiveAR}. In a chip design activity, beginners might receive prompts directing attention to relevant constraints, whereas more experienced students could be asked to diagnose a violation or justify a trade-off before receiving feedback. What constitutes reliable evidence of learner progress in these tasks, however, remains to be determined.

\subsubsection{Using Metaphors to Bridge Intuition and Formal Models}

Some difficult concepts may benefit from temporary spatial analogies. A clock tree could be introduced as a branching network in which relative signal arrival times are visible; multilayer routing could be explored as a connected network of paths and vertical transitions. Such analogies can give students an initial way to reason about relationships that are otherwise presented through dense schematics or reports.

Their limitations are equally important. The transport-network analogy does not capture all the electrical consequences of interconnect geometry, and animated clock pulses are not a substitute for timing analysis. A well-designed lesson would use a metaphor to establish intuition and then explicitly connect it to waveforms, delays, design rules, and standard EDA representations. Otherwise, the simplified explanation may become an obstacle to more precise understanding.

\subsubsection{Extending Access to Fabrication and Collaborative Practice}

Fabrication is difficult to experience directly in most chip design courses. Immersive walkthroughs could connect mask layouts to successive deposition, lithography, and etching steps, allowing students to inspect how features accumulate into a physical device. This could also make the purpose of manufacturing constraints more concrete: a design rule would be linked to a fabrication concern rather than encountered only as an error in a verification report.

Shared environments offer a related opportunity for collaborative design review. Students and instructors could inspect a layout together, annotate a suspected violation, trace a timing path, and compare design alternatives while referring to the same spatial features. The value would come from supporting explanation, critique, and joint decision-making, not simply enabling multiple users to enter the same virtual space.

\begin{figure*}[!t]
\centering

\begin{minipage}{0.23\textwidth}
    \centering
    \includegraphics[width=\linewidth]{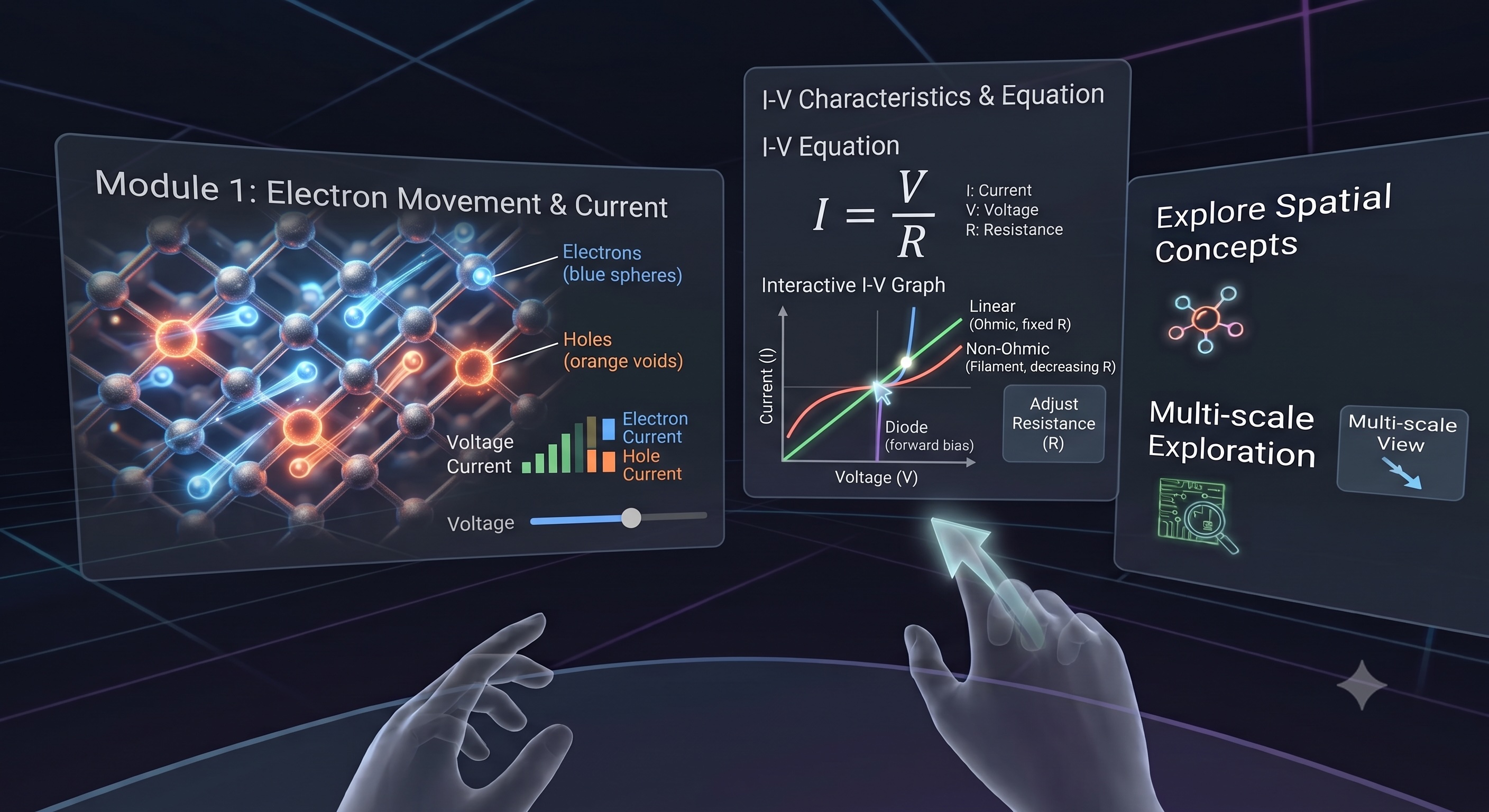}
    
    \small\textbf{(a)}
\end{minipage}
\hfill
\begin{minipage}{0.23\textwidth}
    \centering
    \includegraphics[width=\linewidth]{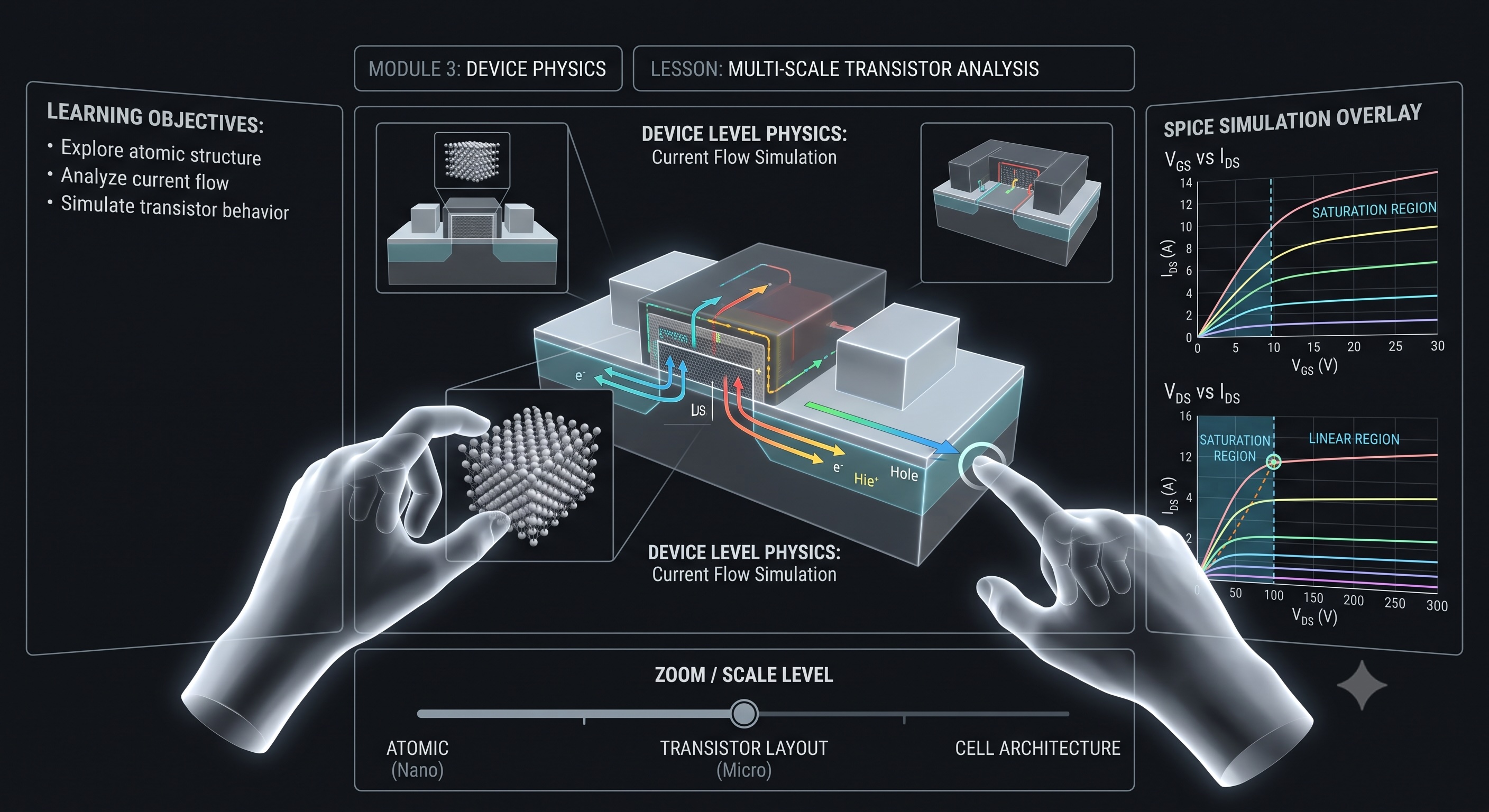}
    
    \small\textbf{(b)}
\end{minipage}
\hfill
\begin{minipage}{0.23\textwidth}
    \centering
    \includegraphics[width=\linewidth]{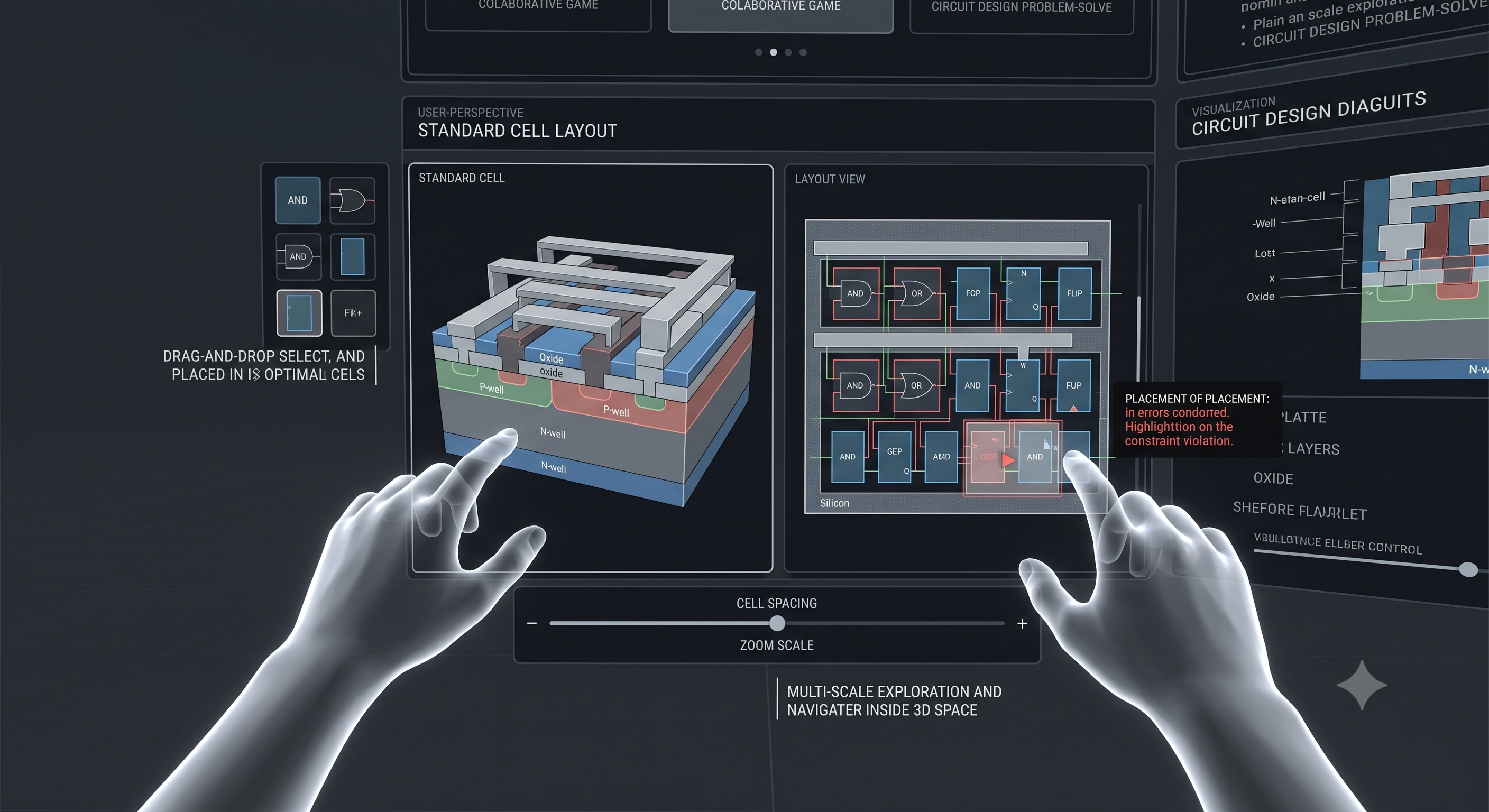}
    
    \small\textbf{(c)}
\end{minipage}
\hfill
\begin{minipage}{0.23\textwidth}
    \centering
    \includegraphics[width=\linewidth]{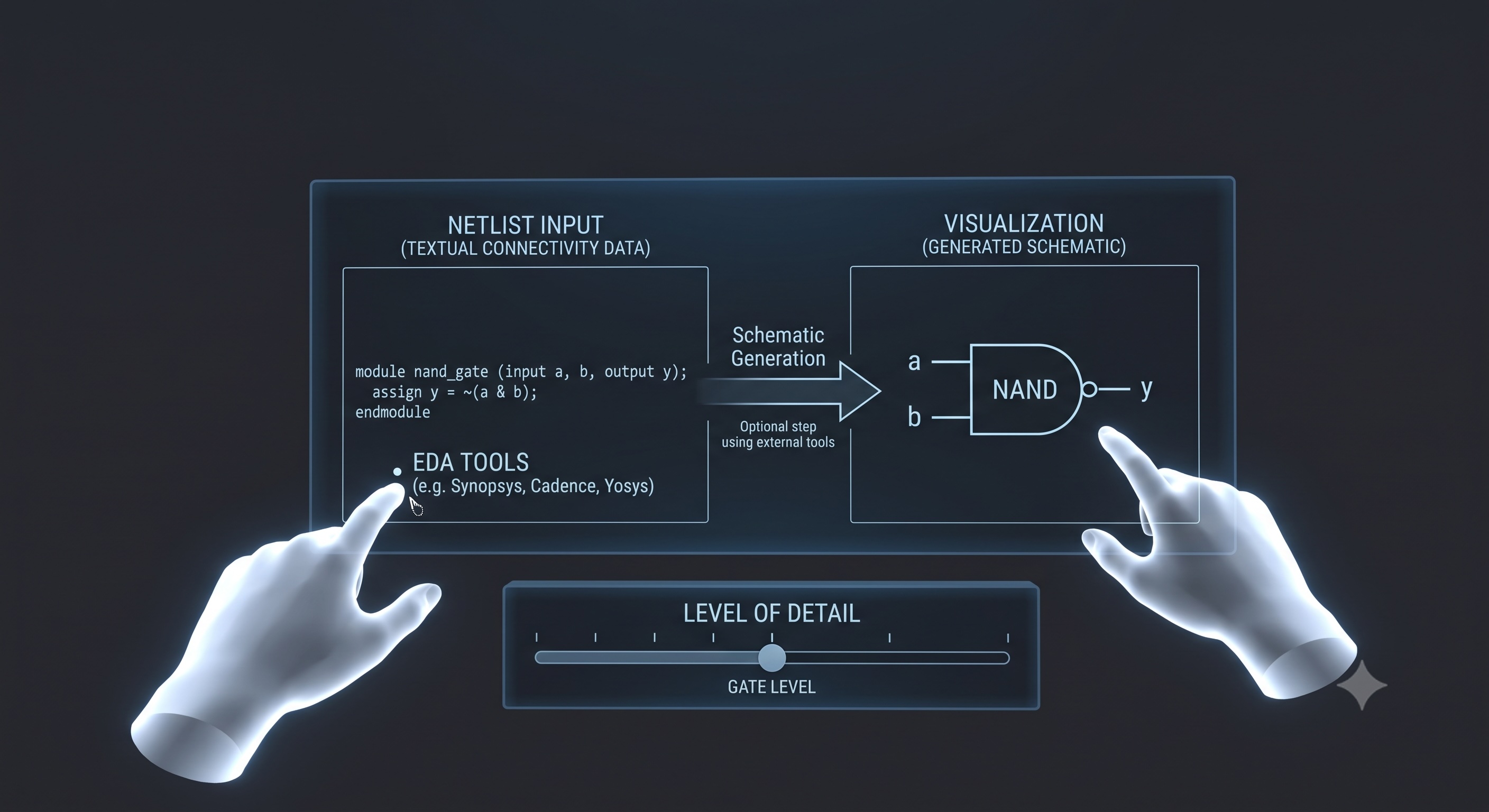}
    
    \small\textbf{(d)}
\end{minipage}

\par\vspace{1em}

\begin{minipage}{0.23\textwidth}
    \centering
    \includegraphics[width=\linewidth]{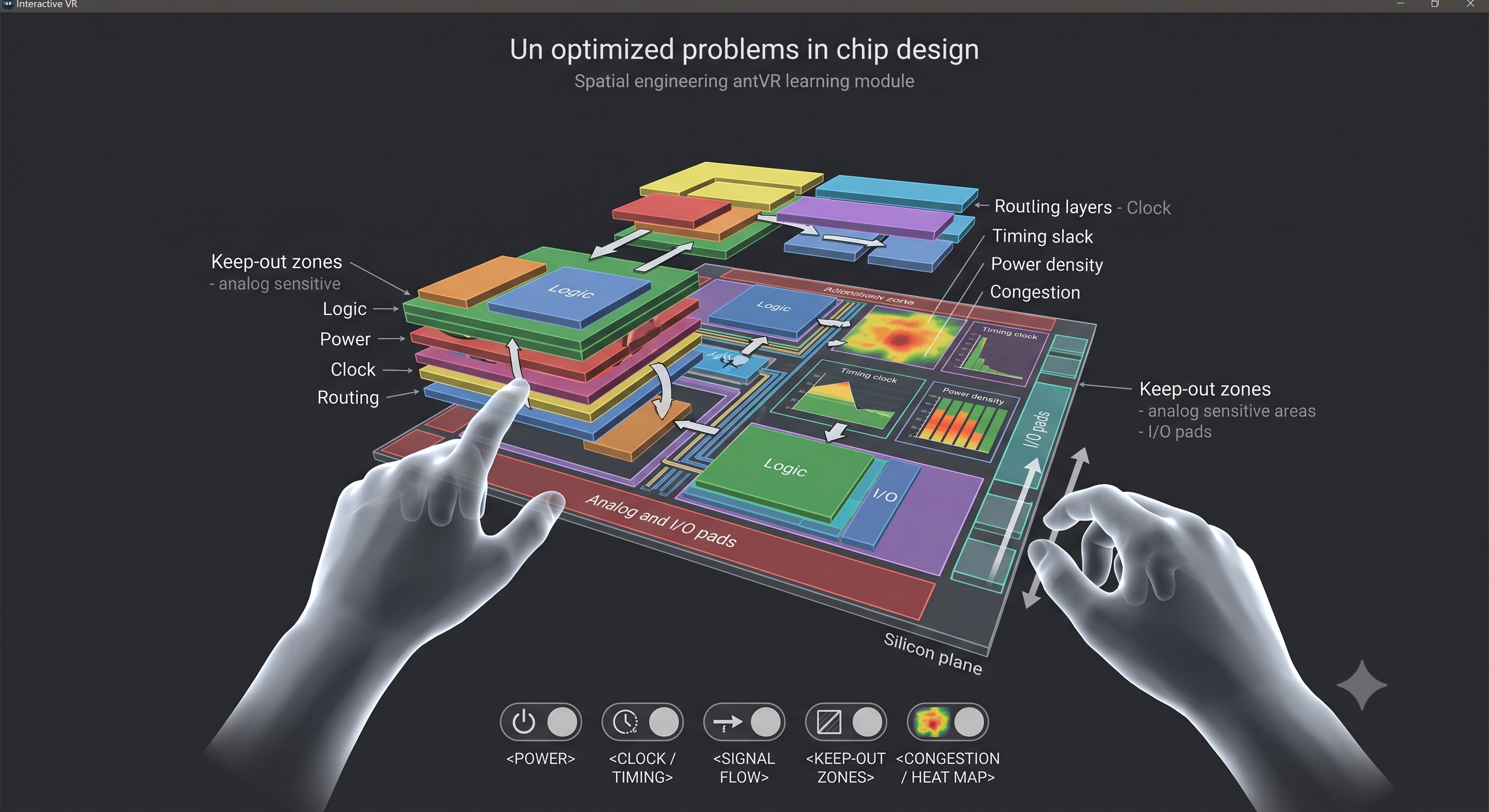}
    
    \small\textbf{(e)}
\end{minipage}
\hfill
\begin{minipage}{0.23\textwidth}
    \centering
    \includegraphics[width=\linewidth]{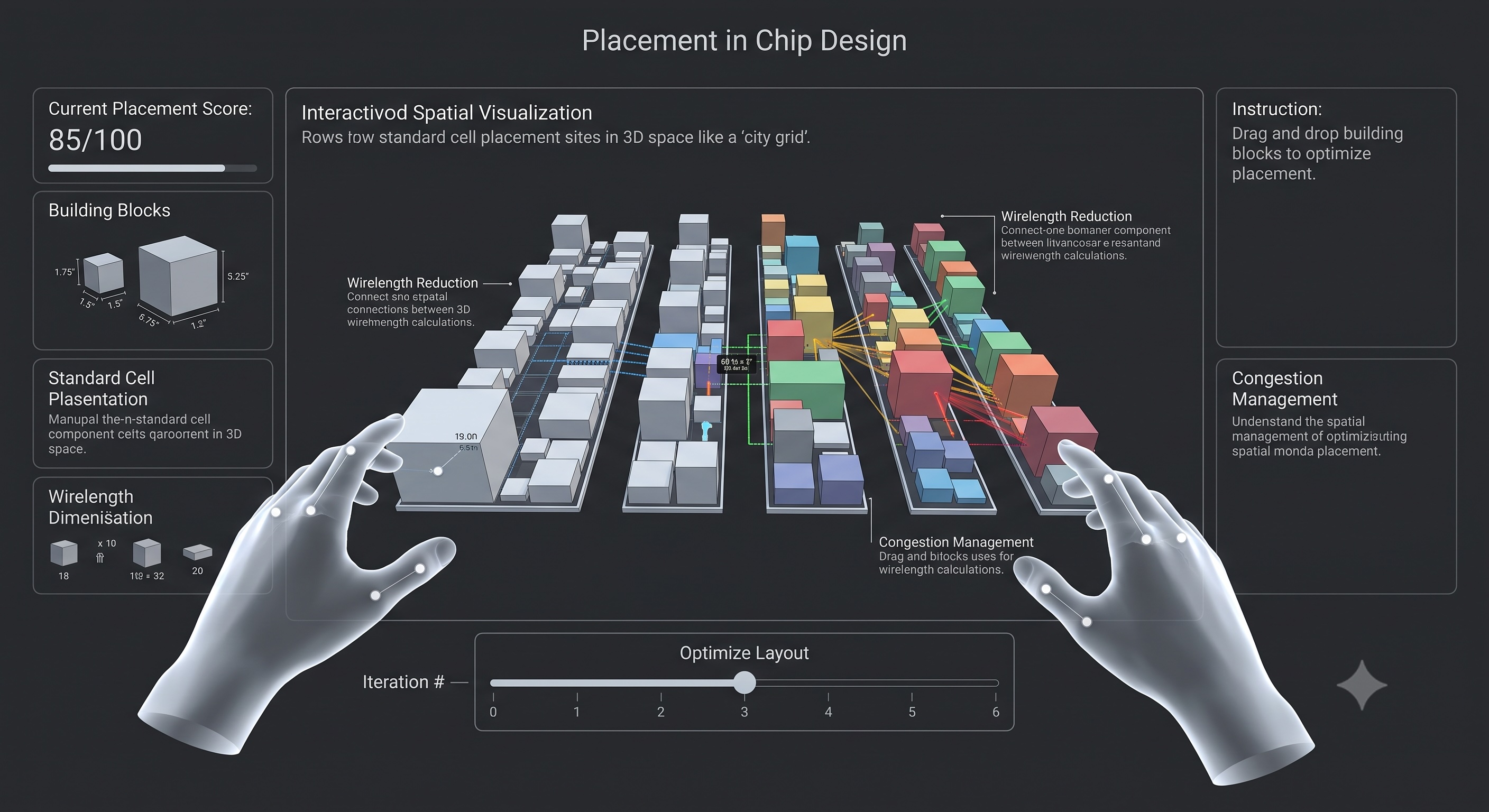}
    
    \small\textbf{(f)}
\end{minipage}
\hfill
\begin{minipage}{0.23\textwidth}
    \centering
    \includegraphics[width=\linewidth]{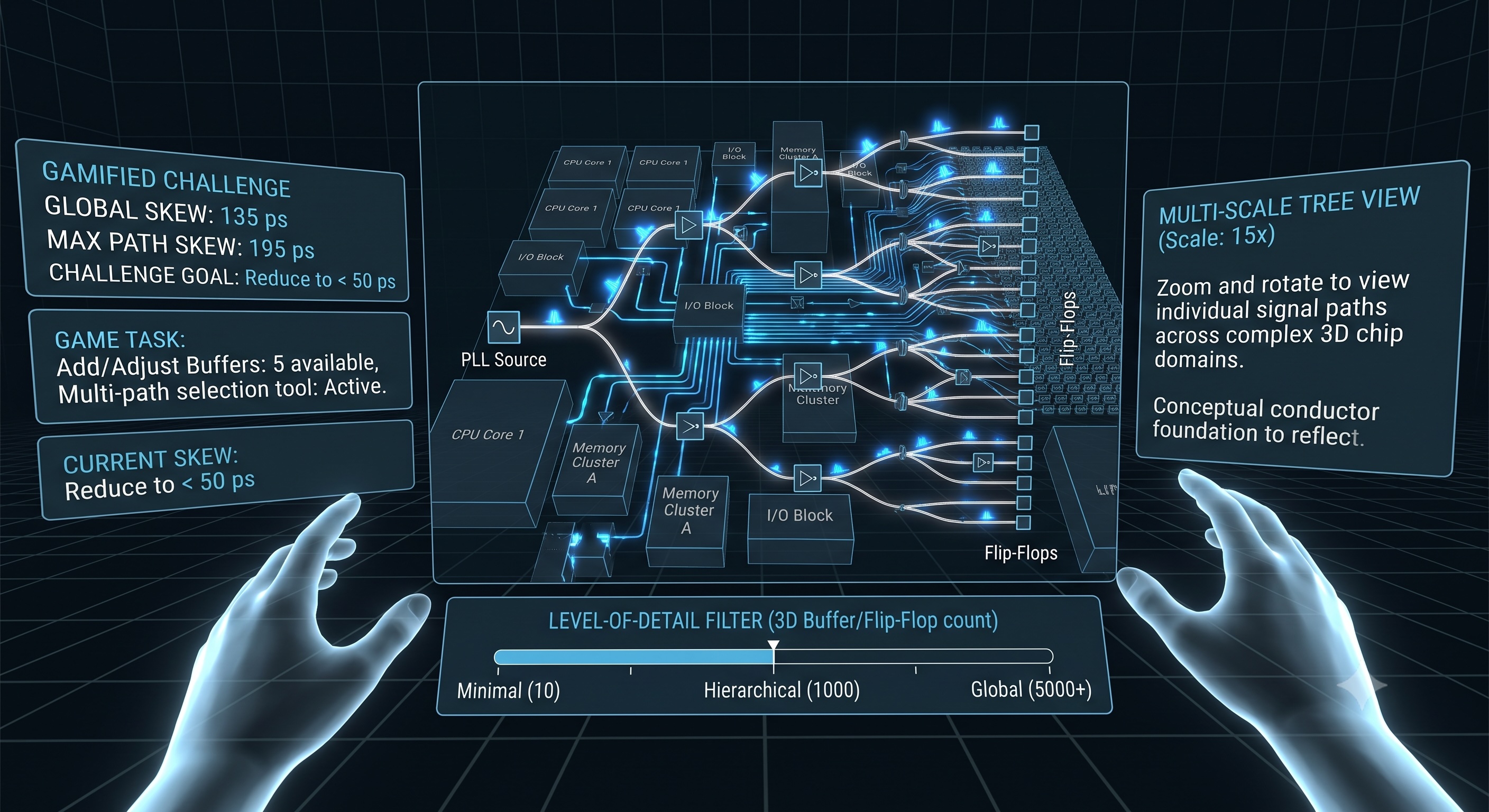}
    
    \small\textbf{(g)}
\end{minipage}
\hfill
\begin{minipage}{0.23\textwidth}
    \centering
    \includegraphics[width=\linewidth]{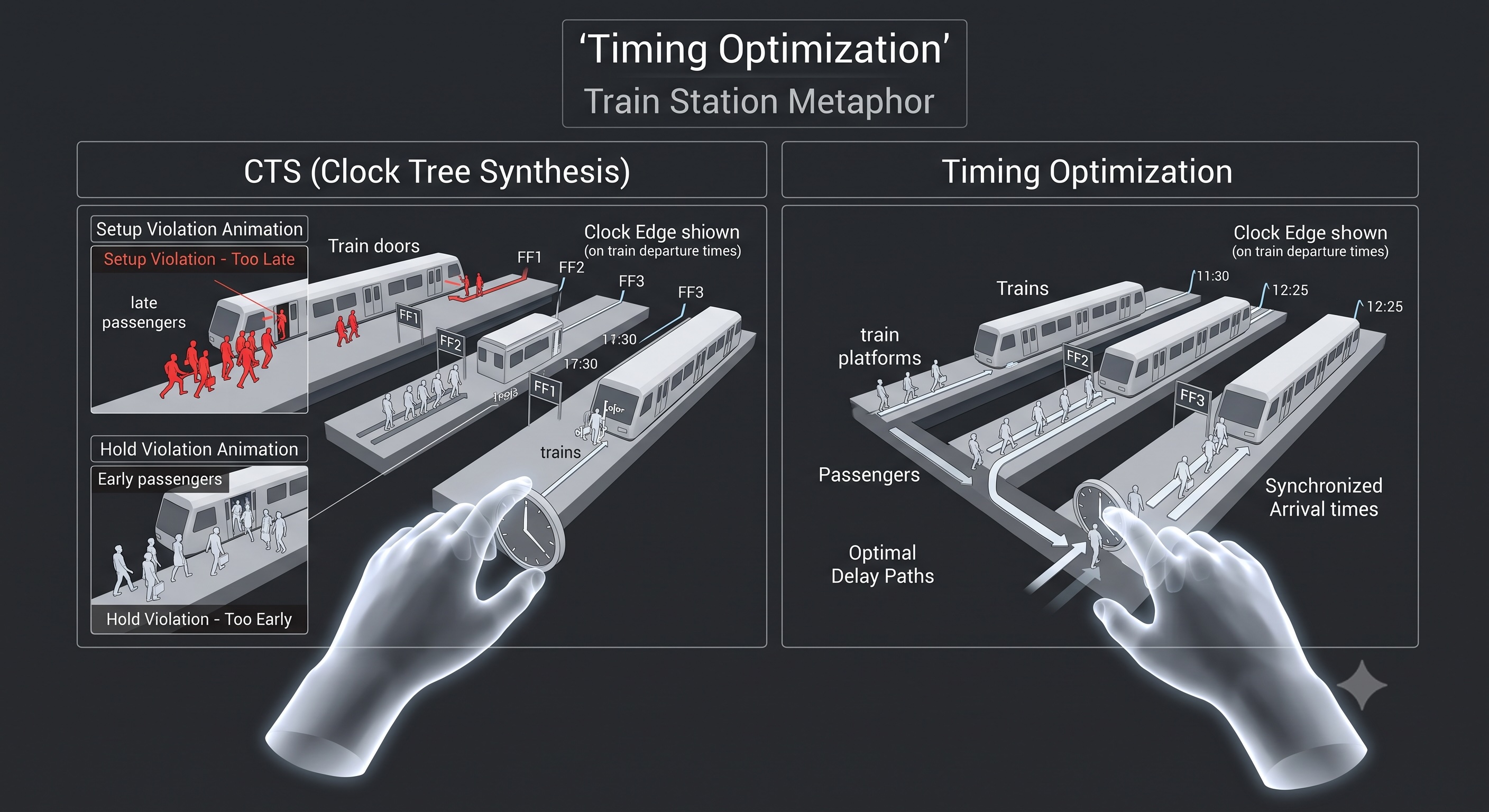}
    
    \small\textbf{(h)}
\end{minipage}

\par\vspace{1em}

\begin{minipage}{0.31\textwidth}
    \centering
    \includegraphics[width=\linewidth]{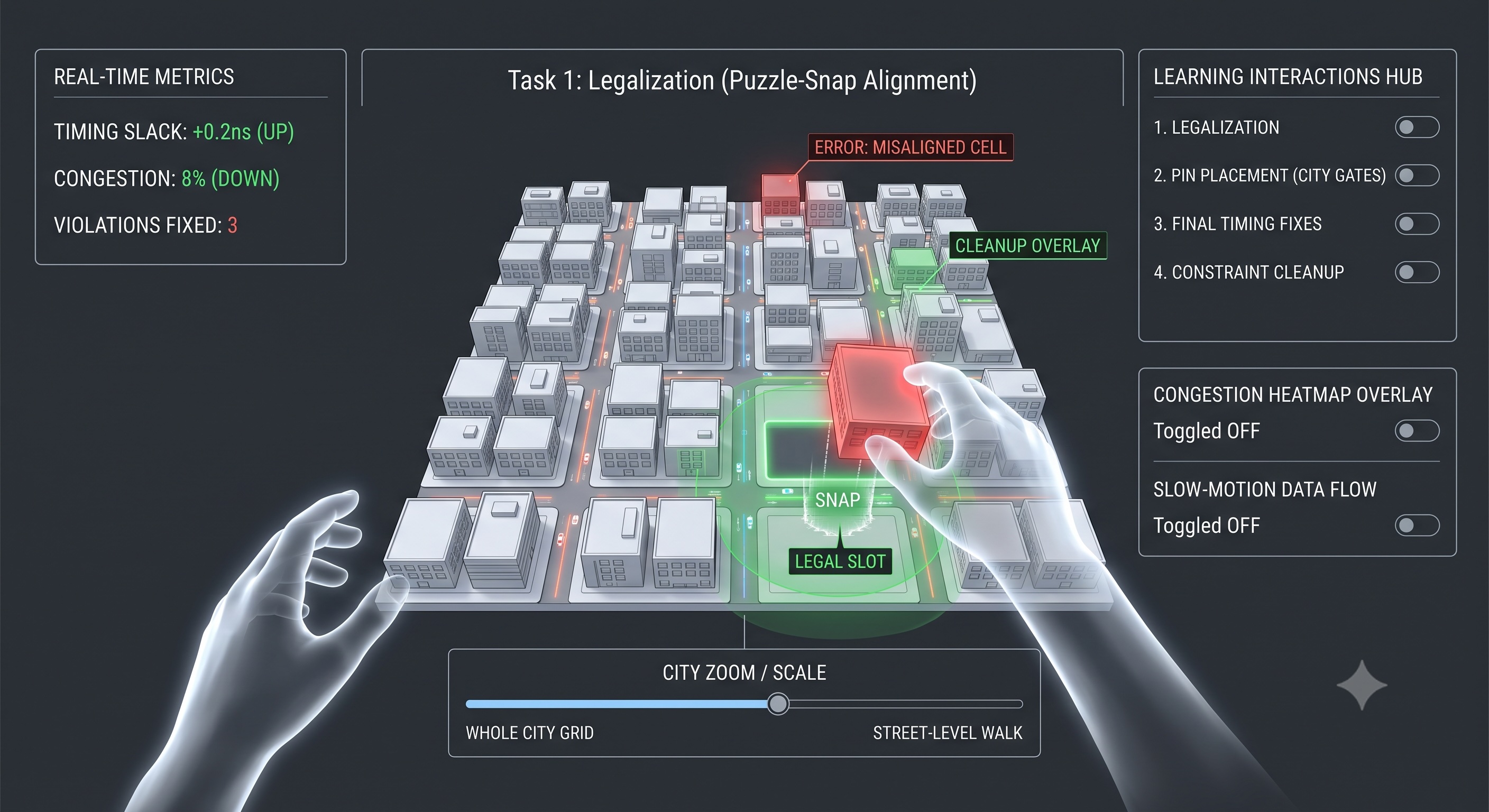}
    
    \small\textbf{(i)}
\end{minipage}
\hfill
\begin{minipage}{0.31\textwidth}
    \centering
    \includegraphics[width=\linewidth]{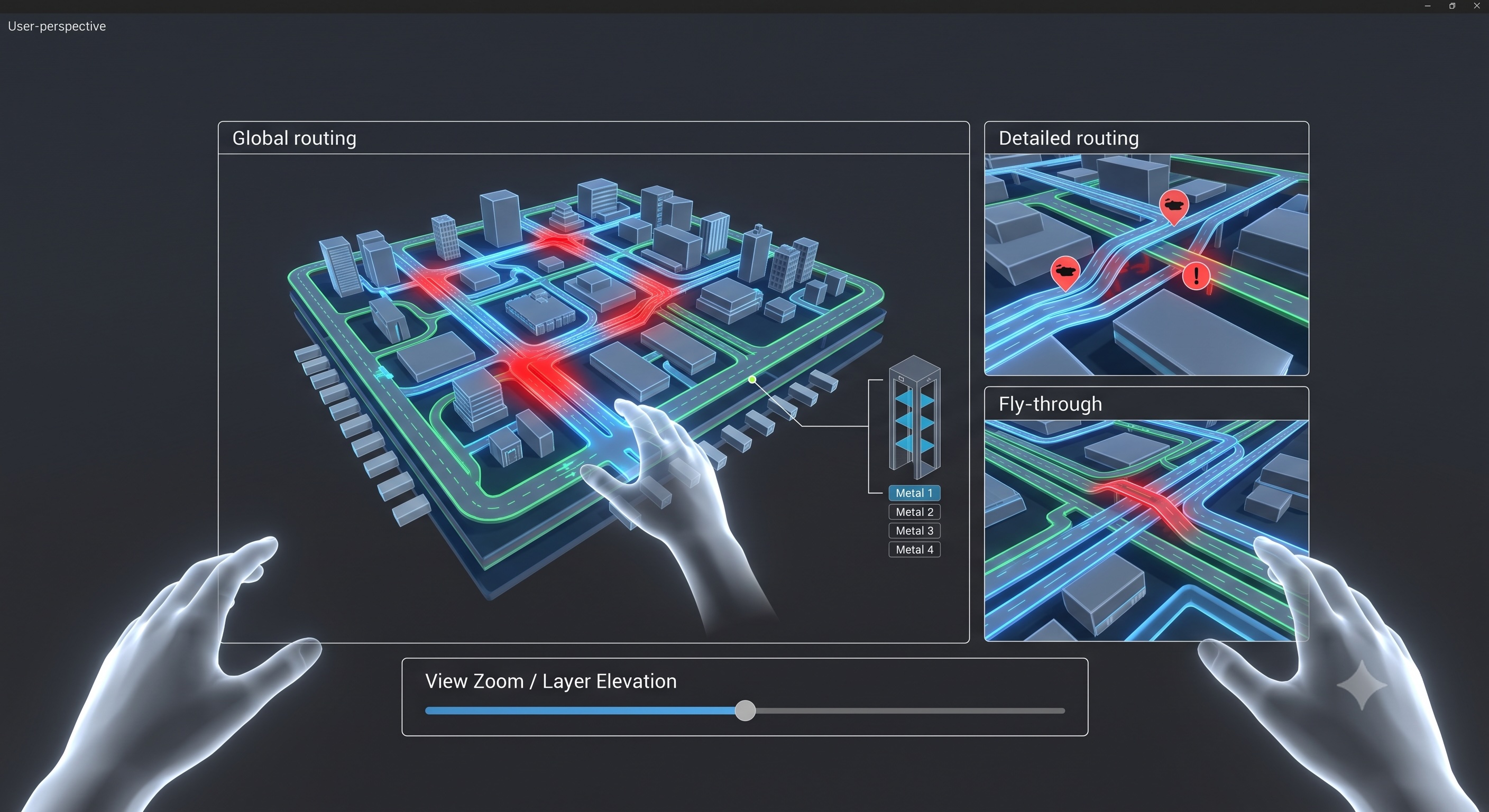}
    
    \small\textbf{(j)}
\end{minipage}
\hfill
\begin{minipage}{0.31\textwidth}
    \centering
    \includegraphics[width=\linewidth]{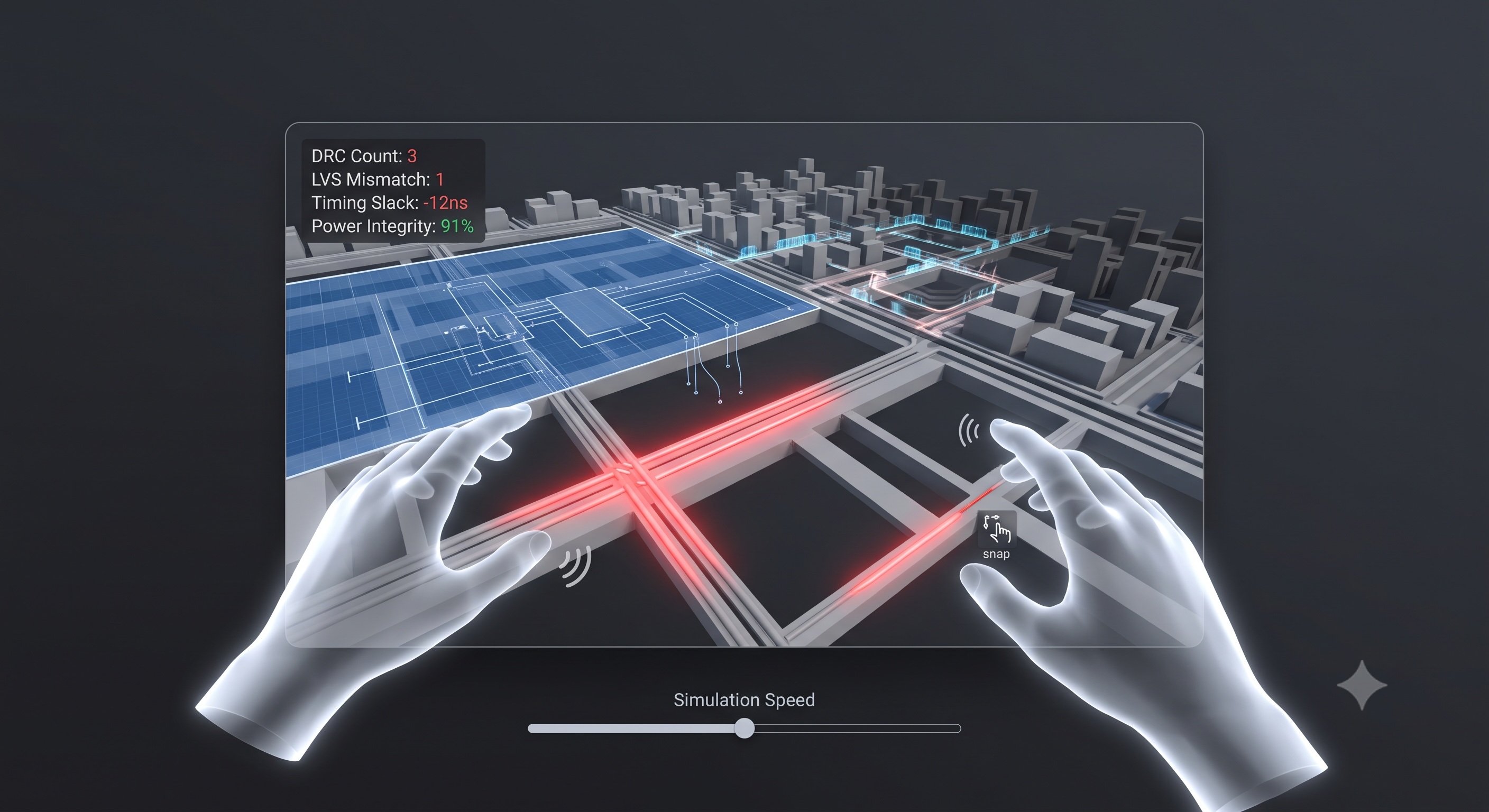}
    
    \small\textbf{(k)}
\end{minipage}

\caption{Illustrative concepts and design stages that may inform
immersive learning activities across the chip design pipeline:
(a) semiconductor physics, (b) transistor structures,
(c) standard cells, (d) logic synthesis,
(e) floorplanning, (f) cell placement,
(g) clock-tree synthesis, (h) timing optimization,
(i) finishing, (j) multilayer routing, and
(k) physical verification. The illustrations provide
context for proposed educational interactions rather than
depicting implemented or evaluated immersive systems.}
\label{fig:chip_design_illustrations}

\end{figure*}

Figure~\ref{fig:chip_design_illustrations} depicts the chip design concepts and stages that motivate these possibilities. It illustrates candidate contexts for learning activities, rather than functioning immersive systems.

\subsubsection{Assessing Reasoning Through Design Processes}

These activities could make aspects of students' reasoning more visible to instructors. A record of design alternatives, revisions, selected views, and responses to feedback might show how a learner approached a constraint and changed their strategy. In floorplanning, for example, the process of recognizing congestion, comparing arrangements, and justifying a final choice may be more informative than the final layout alone.

Action logs cannot explain learner intent on their own. Research on reflection prompts in AR-supported maker learning offers a useful precedent for asking learners to articulate what they attempted and why~\citep{turakhia2023reflectiveAR}. Short design rationales, paired with interaction histories, could help instructors distinguish deliberate reasoning from trial and error. Any resulting assessment method would still need to be validated against independent evidence of understanding.

\subsection{Design and Evaluation Challenges}
\label{sec:discussion-challenges}

Turning these opportunities into effective learning tools requires choices about what to represent, how closely to connect the experience to engineering practice, and what evidence would demonstrate an educational benefit. These issues are particularly consequential in chip design because simplified representations can be useful for novices while also concealing the constraints that determine whether a design is viable.

\subsubsection{Balancing Technical Fidelity and Pedagogical Clarity}

A learning environment need not reproduce an industrial EDA tool, but it must preserve the relationships essential to the learning task. An introductory routing exercise might focus on layers, vias, and connectivity before introducing parasitic delay or manufacturability constraints. Conversely, an exercise about timing closure would be misleading if it allowed learners to optimize geometry without accounting for the electrical and physical effects of their decisions.

This balance calls for deliberate abstraction rather than indiscriminate simplification. Constraints could be introduced progressively as learners gain familiarity with a task, while annotations make clear which effects are modeled and which are omitted. Students should also be able to move between immersive explanations and conventional diagrams or reports, so that understanding does not depend on a particular interface.

\subsubsection{Connecting to Authentic EDA Workflows}

Chip design knowledge is ultimately exercised through HDL, netlists, layout databases, timing reports, and verification outputs. Immersive activities should therefore connect to these artifacts where the learning objective requires it. A layout exploration tool might link selected geometry to its corresponding netlist objects, for example, while a timing activity might overlay path information derived from static timing analysis.

The technical challenges are substantial. Representations must remain consistent across scales, and changes to placement or routing may require expensive re-analysis before accurate feedback can be shown. For some educational tasks, precomputed alternatives or simplified models may be sufficient; others may warrant integration with an open-source design flow. The trade-off should be made explicit rather than suggesting that all consequences can be calculated instantaneously.

\subsubsection{Managing Cognitive Load and Instructional Guidance}

Spatial integration does not guarantee that information becomes easier to understand. Showing timing, congestion, power, and verification overlays simultaneously could make an already demanding task harder, especially for learners unfamiliar with the underlying representations. Navigation and interaction mechanics may also consume attention that should be directed toward the chip design problem.

Activities should therefore focus learners on a small number of consequential relationships at a time, with additional detail revealed as it becomes relevant. Structured comparisons, guided questions, and opportunities to revisit earlier decisions could help learners interpret feedback rather than react to it mechanically. These choices should be informed by learner expertise and the instructional setting, not by a desire to maximize visual complexity.

\subsubsection{Integrating with Existing Instruction}

Immersive activities are more likely to be useful when they address a specific difficulty within a course. A short module on multilayer routing, clock skew, or layout verification may complement lectures and EDA laboratories more effectively than an attempt to recreate an entire chip design course in VR. Such modules also give instructors a clearer basis for defining learning objectives and assessing what students have gained.

Deployment raises practical questions about hardware access, session length, accessibility, maintenance, and collaboration. Instructors need to guide the activity, observe students' progress, and connect the experience to assignments that use authentic engineering representations. These considerations should shape prototype development from the outset rather than being addressed after a system is complete.

\subsubsection{Evaluating Learning and Transfer}

The crucial test is whether immersion changes what students understand and can do. Evaluations should examine learners' ability to connect representations, diagnose timing and routing problems, explain trade-offs, and apply those skills to unfamiliar designs. Measures of engagement or perceived usefulness are insufficient on their own, particularly when the novelty of VR may influence students' responses.

Comparisons with desktop visualization, conventional laboratory exercises, and EDA tools would clarify which outcomes depend on immersion and which follow from better instructional design more generally. Transfer to non-immersive tools deserves particular attention, since students must ultimately work with the representations and workflows used in chip design practice. These comparisons can help determine not only whether an immersive intervention works, but where its additional complexity is justified.

Overall, the findings support a selective, learning-centered approach to immersive chip design education. The strongest opportunities are tasks in which learners must connect representations, interpret relationships that are difficult to see, and reason about the consequences of decisions. Designing for those tasks---and evaluating the resulting understanding---offers a more defensible direction than treating immersion as an educational benefit in itself.

\section{Limitations and Future Work}
\label{sec:limitations}

This study offers an exploratory account of learning challenges in chip design education and identifies opportunities for immersive technologies to address them. Several methodological and conceptual limitations shape the interpretation of our findings and point toward directions for future research.

First, our analysis draws on interviews with five participants whose experiences span chip design instruction, teaching support, and advanced coursework. Although their perspectives provide insight into curricular demands and instructional challenges, the small sample and shared institutional context limit the transferability of our findings. Future research should examine these challenges across a wider range of institutions, educational levels, and instructional settings, including undergraduate and graduate programs, community colleges, and industry training environments. Such work could reveal how learning demands vary with curricular emphasis, prior experience, and access to design tools.

Second, our findings are informed by curriculum analysis and qualitative interviews rather than direct observations or assessments of student learning. The identified challenges therefore reflect curricular requirements and participants' reported experiences, rather than independently measured learning difficulties or outcomes. Classroom observations, learner interviews, analysis of design artifacts, and performance-based assessments could provide a more detailed understanding of how students navigate abstraction levels, interpret spatial and temporal relationships, and reason about design constraints. These investigations would also help refine Cross-Scale Spatial Reasoning (CSSR), including its constituent dimensions, relationship to established reasoning constructs, and potential measures of learner development.

Third, the mapping between learning demands and immersive affordances represents a conceptual synthesis rather than an empirically validated instructional framework. While it identifies opportunities for VR and AR, the analysis does not establish whether immersive approaches improve learning relative to existing methods. Future work should translate selected opportunities into targeted learning experiences, particularly for tasks involving device structures, multilayer routing, timing analysis, and design optimization. Evaluations combining controlled studies with classroom deployments could then examine how different forms of visualization, interaction, and feedback support learning, and under what conditions they offer advantages over conventional diagrams, desktop simulations, or EDA tools.

Finally, developing effective immersive learning environments will require careful attention to technical fidelity, instructional integration, and learning assessment. Connecting immersive representations to authentic design artifacts and simulation outputs could make learning activities more meaningful, but also introduces challenges related to data interoperability, computational complexity, and the representation of engineering constraints. Equally important is determining whether these environments improve students' ability to understand concepts, interpret relationships, and make informed design decisions. Beyond usability and engagement, evaluations should examine reasoning performance and transfer to conventional design workflows. Process-based evidence, such as learners' design choices, revisions, and responses to feedback, may offer further insight into how understanding develops, provided these measures are validated against meaningful learning outcomes.

Together, these directions establish a path from the exploratory findings and conceptual mapping presented here toward the design and empirical evaluation of immersive learning environments. The broader goal is to determine when, how, and for whom immersive technologies can meaningfully support reasoning across physical scales, design representations, and interconnected decisions in chip design education.

\section{Conclusion}
\label{sec:conclusion}

Learning chip design requires more than mastering individual concepts and workflows. Students must connect physical structures, design representations, and engineering decisions across multiple scales and stages of the design pipeline. These relationships can be difficult to interpret through conventional instructional materials, particularly when learners must reason about spatial configurations, temporal behavior, and the downstream consequences of design choices. Through curriculum analysis and qualitative interviews with five participants, this study examined these interconnected learning demands and the opportunities they present for immersive technologies.

Our analysis characterized four categories of learning demands---descriptive knowledge, procedural knowledge, relational and design-focused reasoning, and strategic decision-making---and identified recurring challenges in connecting abstraction levels, interpreting spatial structures, understanding temporal behavior, applying theoretical knowledge, and reasoning about design trade-offs. We subsequently developed a conceptual mapping between these learning demands and six immersive affordances, identifying where targeted visualization, interaction, and feedback may support learning across the chip design pipeline.

These findings motivate Cross-Scale Spatial Reasoning (CSSR) as a conceptual lens for understanding how learners connect physical structures, design representations, and the consequences of engineering decisions. More broadly, they suggest that the educational potential of immersive technologies lies not in reproducing chip design workflows in VR, but in making relationships that are otherwise difficult to interpret more accessible to learners. By grounding immersive design opportunities in the reasoning demands of chip design, this work provides a foundation for developing and evaluating learning environments that complement existing instruction and foster a more integrated understanding of semiconductor design.

\vspace{6pt}

\reftitle{References}


\bibliography{references}

@misc{zerotoasiccourse,
  author       = {{Zero to ASIC Course}},
  title        = {{Zero to ASIC Course}},
  year         = {2026},
  url          = {https://www.zerotoasiccourse.com/},
}

@article{venn2024tinytapeout,
  author  = {Venn, Matthew David},
  title   = {{Tiny Tapeout}: A Shared Silicon Tapeout Platform Accessible to Everyone},
  journal = {IEEE Solid-State Circuits Magazine},
  volume  = {16},
  number  = {2},
  pages   = {20--29},
  year    = {2024},
  doi     = {10.1109/MSSC.2024.3381097}
}

@article{zelioli2026croc,
  title   = {Croc: Training the Next Generation Chip Designers on Domain-Specific End-to-End Open Source Silicon},
  author  = {Zelioli, Enrico and Sauter, Philippe and Benz, Thomas and Pochert, Hannah and W{\"u}thrich, Luisa and Muheim, Beat and G{\"u}rkaynak, Frank K. and Benini, Luca},
  journal = {arXiv preprint arXiv:2606.25673},
  year    = {2026}
}

@article{krupp2026rtlprompt,
  title   = {From RTL to Prompt Coding: Empowering the Next Generation of Chip Designers through LLMs},
  author  = {Krupp, Lukas and Venn, Matthew and Wehn, Norbert},
  journal = {arXiv preprint arXiv:2601.13815},
  year    = {2026}
}

@book{dewey1938experience,
  author    = {Dewey, John},
  title     = {Experience and Education},
  publisher = {Macmillan},
  year      = {1938}
}

@book{kolb1984experiential,
  author    = {Kolb, David A.},
  title     = {Experiential Learning: Experience as the Source of Learning and Development},
  publisher = {Prentice Hall},
  year      = {1984}
}

@article{dalgarno2010learning,
  author  = {Dalgarno, Barney and Lee, Mark J. W.},
  title   = {What are the learning affordances of 3-D virtual environments?},
  journal = {British Journal of Educational Technology},
  volume  = {41},
  number  = {1},
  pages   = {10--32},
  year    = {2010},
  doi     = {10.1111/j.1467-8535.2009.01038.x}
}

@article{merchant2014effectiveness,
  author  = {Merchant, Zahira and Goetz, Ernest T. and Cifuentes, Lauren and Keeney-Kennicutt, Wendy and Davis, Trina J.},
  title   = {Effectiveness of virtual reality-based instruction on students' learning outcomes in K-12 and higher education: A meta-analysis},
  journal = {Computers \& Education},
  volume  = {70},
  pages   = {29--40},
  year    = {2014},
  doi     = {10.1016/j.compedu.2013.07.033}
}

@article{radianti2020review,
  author  = {Radianti, Jaziar and Majchrzak, Tim A. and Fromm, Jennifer and Wohlgenannt, Isabell},
  title   = {A systematic review of immersive virtual reality applications for higher education: Design elements, lessons learned, and research agenda},
  journal = {Computers \& Education},
  volume  = {147},
  pages   = {103778},
  year    = {2020},
  doi     = {10.1016/j.compedu.2019.103778}
}

@misc{ref-url1-deloitte,
  author       = {{Deloitte Study}},
  title        = {Semiconductor Industry Outlook},
  howpublished = {Available online},
  url          = {https://www.deloitte.com/us/en/insights/industry/technology/technology-media-telecom-outlooks/semiconductor-industry-outlook.html},
  year         = 2026
}

@misc{ref-SEMI-global-talent-gap,
  author       = {{SEMI}},
  title        = {The Semiconductor Talent Crisis: Why Growing Demand Can't Find Leaders},
  howpublished = {Available online},
  url          = {https://www.semi.org/en/blogs/the-semiconductor-talent-crisis-why-growing-demand-cant-find-leaders},
  year         = 2025
}

@misc{ref-url3-talent-gap,
  author       = {{Sourceability Team}},
  title        = {Overcoming Semiconductor Talent Gaps + Supply Challenges},
  howpublished = {Available online},
  url          = {https://sourceability.com/post/overcoming-semiconductor-talent-challenges},
  year         = 2025
}

@article{ref-journal4-chips-act,
  author  = {Damanpak Rizi, A. and Roy, A. and Noor, R. and Kang, H. and Varshney, N. and Jacob, K. and Rivera-Jimenez, S. and Edwards, N. and Sorger, V. J. and Dalir, H. and others},
  title   = {From Talent Shortage to Workforce Excellence in the {CHIPS} Act Era: Harnessing Industry 4.0 Paradigms for a Sustainable Future in Domestic Chip Production},
  journal = {arXiv},
  year    = {2023},
}

@article{ref-journal7,
  author  = {Dickerson, S. J. and Clark, R. M.},
  title   = {A Classroom-Based Simulation-Centric Approach to Microelectronics Education},
  journal = {Computer Applications in Engineering Education},
  year    = {2018},
  volume  = {26},
  pages   = {768--781}
}

@article{ref-journal9-razavi2024education,
  author  = {Razavi, B.},
  title   = {Education of Chip Designers at a Large Scale: A Proposal [Society News]},
  journal = {IEEE Solid-State Circuits Magazine},
  year    = {2024},
  volume  = {16},
  pages   = {76--83}
}

@article{ref-journal10-gulgonul2025sparking,
  author  = {Gulgonul, S.},
  title   = {Sparking Curiosity in Digital System Design Lectures with Take Home Labs},
  journal = {arXiv},
  year    = {2025},
  note    = {arXiv:2503.16625}
}

@inproceedings{ref-proceeding11-dunn2023hands,
  author    = {Dunn, S. and W{\"u}nsche, B. C. and Allison, J. R. and Thompson, S. E. R. and Lange-Nawka, D.},
  title     = {Hands-on {DNA}: Exploring the Impact of Virtual Reality on Teaching {DNA} Structure and Function},
  booktitle = {Proceedings of the 29th ACM Symposium on Virtual Reality Software and Technology},
  year      = {2023},
  pages     = {1--11}
}

@inproceedings{ref-proceeding12-amaral2024interactive,
  author    = {Amaral, L. and Santos, J. and Soares, F. and Felix, J. and Cardoso, L. and Nascimento, T. H.},
  title     = {Interactive Educational Game for Chemistry in Virtual Reality with User Gesture Interactions via Smartwatches},
  booktitle = {Proceedings of the 26th Symposium on Virtual and Augmented Reality},
  year      = {2024},
  pages     = {294--298}
}

@article{ref-journal13-strzys2018physics,
  author  = {Strzys, M. P. and Kapp, S. and Thees, M. and Klein, P. and Lukowicz, P. and Knierim, P. and Schmidt, A. and Kuhn, J.},
  title   = {Physics Holo.lab Learning Experience: Using Smartglasses for Augmented Reality Labwork to Foster the Concepts of Heat Conduction},
  journal = {European Journal of Physics},
  year    = {2018},
  volume  = {39},
  pages   = {035703}
}

@article{ref-journal14-warkentin2025hands,
  author  = {Warkentin, M. and Altmeyer, K. and Liang, Y. and Steinmacher, B. and Gr{\"a}nz, B. and Lichtenberger, A. and K{\"u}chemann, S. and Kuhn, J. and Hoyer, C.},
  title   = {Hands-on Experiment Supported by Augmented Reality Smartglasses for Learning the Lorentz Force},
  journal = {arXiv},
  year    = {2025},
  note    = {arXiv:2502.06421}
}

@article{ref-journal15-ka2025analysis,
  author  = {Ka, J. and Kim, H. and Kim, J. and Kim, W.},
  title   = {Analysis of Virtual Reality Teaching Methods in Engineering Education: Assessing Educational Effectiveness and Understanding of 3D Structures},
  journal = {Virtual Reality},
  year    = {2025},
  volume  = {29},
  pages   = {17}
}

@article{ref-journal16-paszkiewicz2021vr,
  author  = {Paszkiewicz, A. and Salach, M. and Strza{\l}ka, D. and Budzik, G. and Nikodem, A. and W{\'o}jcik, H. and Witek, M.},
  title   = {{VR} Education Support System: A Case Study of Digital Circuits Design},
  journal = {Energies},
  year    = {2021},
  volume  = {15},
  pages   = {277}
}

@article{ref-journal17-vidak2024augmented,
  author  = {Vidak, A. and {\v{S}}api{\'c}, I. M. and Me{\v{s}}i{\'c}, V. and Gomzi, V.},
  title   = {Augmented Reality Technology in Teaching About Physics: A Systematic Review of Opportunities and Challenges},
  journal = {European Journal of Physics},
  year    = {2024},
  volume  = {45},
  pages   = {023002}
}

@article{ref-journal19-naeem2023step,
  author  = {Naeem, M. and Ozuem, W. and Howell, K. and Ranfagni, S.},
  title   = {A Step-by-Step Process of Thematic Analysis to Develop a Conceptual Model in Qualitative Research},
  journal = {International Journal of Qualitative Methods},
  year    = {2023},
  volume  = {22},
  pages   = {16094069231205789}
}

@book{ref-book20-anderson2001taxonomy,
  author    = {Anderson, L. W. and Krathwohl, D. R.},
  title     = {A Taxonomy for Learning, Teaching, and Assessing: A Revision of Bloom's Taxonomy of Educational Objectives: Complete Edition},
  publisher = {Addison Wesley Longman, Inc.},
  year      = {2001}
}

@book{ref-book21-cross2006designerly,
  author    = {Cross, N.},
  title     = {Designerly Ways of Knowing},
  publisher = {Springer},
  year      = {2006}
}

@article{ref-journal23-parong2018learning,
  author  = {Parong, J. and Mayer, R. E.},
  title   = {Learning Science in Immersive Virtual Reality},
  journal = {Journal of Educational Psychology},
  year    = {2018},
  volume  = {110},
  pages   = {785}
}

@article{ref-journal24-bailenson2008use,
  author  = {Bailenson, J. N. and Yee, N. and Blascovich, J. and Beall, A. C. and Lundblad, N. and Jin, M.},
  title   = {The Use of Immersive Virtual Reality in the Learning Sciences: Digital Transformations of Teachers, Students, and Social Context},
  journal = {Journal of the Learning Sciences},
  year    = {2008},
  volume  = {17},
  pages   = {102--141}
}

@article{ref-journal25-ragan2010effects,
  author  = {Ragan, E. D. and Sowndararajan, A. and Kopper, R. and Bowman, D. A.},
  title   = {The Effects of Higher Levels of Immersion on Procedure Memorization Performance and Implications for Educational Virtual Environments},
  journal = {Presence},
  year    = {2010},
  volume  = {19},
  pages   = {527--543}
}

@misc{ref-url26-ucscsiliconvalleyext,
  author  = {{UC Santa Cruz Extension}},
  title        = {Learn More: Chip Design \& Semiconductor Engineering},
  howpublished = {Available online},
  url          = {https://www.ucsc-extension.edu/learn-more-chip-design-semiconductor-engineering/},
}

@article{ref-journal27-abhishek2023effectiveness,
  author  = {Abhishek, N. and Kulal, A. and Divyashree, M. S. and Dinesh, S.},
  title   = {Effectiveness of {MOOCs} on Learning Efficiency of Students: A Perception Study},
  journal = {Journal of Research in Innovative Teaching \& Learning},
  year    = {2023},
  volume  = {18},
  pages   = {145--164}
}

@article{ref-journal32-yang2025nls,
  author  = {Yang, K. and Ouyang, H. and Wang, X. and Lu, B. and Wang, Y. and Abhayaratne, C. and Li, S. and Jin, L. and Deng, T.},
  title   = {{NLS}: Natural-Level Synthesis for Hardware Implementation Through {GenAI}},
  journal = {arXiv},
  year    = {2025},
  note    = {arXiv:2504.01981}
}

@article{ref-journal35-abdollahi2025hardware,
  author  = {Abdollahi, M. and Yeganli, S. F. and Baharloo, M. and Baniasadi, A.},
  title   = {Hardware Design and Verification with Large Language Models: A Scoping Review, Challenges, and Open Issues},
  journal = {Electronics},
  year    = {2025},
  volume  = {14},
  pages   = {120}
}

@article{ref-journal36-klingenberg2024embodiment,
  author  = {Klingenberg, S. and Bosse, R. and Mayer, R. E. and Makransky, G.},
  title   = {Does Embodiment in Virtual Reality Boost Learning Transfer? Testing an Immersion-Interactivity Framework},
  journal = {Educational Psychology Review},
  year    = {2024},
  volume  = {36},
  pages   = {116}
}

@article{ref-journal37-huang2025extended,
  author  = {Huang, T.-C. and Tseng, H.-P.},
  title   = {Extended Reality in Applied Sciences Education: A Systematic Review},
  journal = {Applied Sciences},
  year    = {2025},
  volume  = {15},
  pages   = {4038}
}

@inproceedings{ref-proceeding38-ipsita2022vrwelding,
  author    = {Ipsita, A. and Erickson, L. and Dong, Y. and Huang, J. and Bushinski, A. K. and Saradhi, S. and Villanueva, A. M. and Peppler, K. A. and Redick, T. S. and Ramani, K.},
  title     = {Towards Modeling of Virtual Reality Welding Simulators to Promote Accessible and Scalable Training},
  booktitle = {Proceedings of the CHI Conference on Human Factors in Computing Systems},
  year      = {2022},
  pages     = {1--21}
}

@inproceedings{ref-proceeding39-henstrom2023immersive,
  author    = {Henstrom, J. and De Amicis, R. and Sanchez, C. A. and Turkan, Y.},
  title     = {Immersive Learning in Engineering: A Comparative Study of {VR} and Traditional Building Inspection Methods},
  booktitle = {Proceedings of the 28th International ACM Conference on 3D Web Technology},
  year      = {2023},
  pages     = {1--11}
}

@inproceedings{ref-proceeding40-lee2025immersive,
  author    = {Lee, J. H. and Sutthiwanna, S. and Shin, S. and Jin, T. and Myung, H. and Bae, S.-H.},
  title     = {Immersive Prototyping for Robot Design with 3D Sketching and {VR} Acting in Reconstructed Workspace},
  booktitle = {Proceedings of the Extended Abstracts of the CHI Conference on Human Factors in Computing Systems},
  year      = {2025},
  pages     = {1--7}
}

@inproceedings{ref-proceeding41-zhu2023learniotvr,
  author    = {Zhu, Z. and Liu, Z. and Zhang, Y. and Zhu, L. and Huang, J. and Villanueva, A. M. and Qian, X. and Peppler, K. and Ramani, K.},
  title     = {{Learniotvr}: An End-to-End Virtual Reality Environment Providing Authentic Learning Experiences for Internet of Things},
  booktitle = {Proceedings of the CHI Conference on Human Factors in Computing Systems},
  year      = {2023},
  pages     = {1--17}
}

@article{ainsworth2006deft,
  author  = {Ainsworth, Shaaron},
  title   = {{DeFT}: A Conceptual Framework for Considering
             Learning with Multiple Representations},
  journal = {Learning and Instruction},
  volume  = {16},
  number  = {3},
  pages   = {183--198},
  year    = {2006},
  doi     = {10.1016/j.learninstruc.2006.03.001}
}

@article{makransky2019presence,
  author  = {Makransky, Guido and Terkildsen,
             Thomas Schj{\o}dt and Mayer, Richard E.},
  title   = {Adding Immersive Virtual Reality to a
             Science Lab Simulation Causes More
             Presence but Less Learning},
  journal = {Learning and Instruction},
  volume  = {60},
  pages   = {225--236},
  year    = {2019},
  doi     = {10.1016/j.learninstruc.2017.12.007}
}

@inproceedings{lie2023drilling,
  author = {Lie, Hing and Studer, Kachina and Zhao, Zhen and Thomson, Ben and Turakhia, Dishita G. and Liu, John},
  title = {Training for Open-Ended Drilling through a Virtual Reality Simulation},
  booktitle = {2023 IEEE International Symposium on Mixed and Augmented Reality (ISMAR)},
  year = {2023},
  publisher = {IEEE},
  doi = {10.1109/ISMAR59233.2023.00051}
}

@inproceedings{studer2024openended,
  author = {Studer, Kachina and Lie, Hing and Zhao, Zhen and Thomson, Ben and Turakhia, Dishita G. and Liu, John},
  title = {An Open-ended System in Virtual Reality for Training Machining Skills},
  booktitle = {Extended Abstracts of the CHI Conference on Human Factors in Computing Systems},
  year = {2024},
  publisher = {ACM},
  doi = {10.1145/3613905.3648666}
}

@inproceedings{patel2024benchmarking,
  author = {Patel, Aarsh and Turakhia, Dishita and Brunner, Oliver and Mertens, Alexander and Liu, John},
  title = {Benchmarking Open-endedness of VR Training Systems in Manufacturing},
  booktitle = {2024 IEEE International Symposium on Mixed and Augmented Reality Adjunct (ISMAR-Adjunct)},
  pages = {64--70},
  year = {2024},
  publisher = {IEEE},
  doi = {10.1109/ISMAR-Adjunct64951.2024.00024}
}

@inproceedings{sepulveda2024actr,
  author = {Sepulveda, Francisco and Fan, Yibo and Gabbianelli, Christian and Studer, Kachina and Turakhia, Dishita G. and Liu, John},
  title = {Using {ACT-R} Architecture in the Design of Intelligent Tutoring Systems for {VR} Training of Manufacturing Skills},
  booktitle = {2024 IEEE International Symposium on Mixed and Augmented Reality Adjunct (ISMAR-Adjunct)},
  year = {2024},
  publisher = {IEEE},
  doi = {10.1109/ISMAR-Adjunct64951.2024.00028}
}

@inproceedings{turakhia2025adaptiveAR,
  author = {Turakhia, Dishita G. and Parent, Mark and Grossman, Tovi and Glueck, Michael and Lafreniere, Ben},
  title = {Investigating Augmented Reality for Adaptive Motor-Skill Training},
  booktitle = {The 51st Graphics Interface Conference 2025},
  year = {2025},
  articleno = {19},
  pages = {1--10},
  publisher = {ACM},
  doi = {10.1145/3769872.3769891}
}

@article{turakhia2024educators,
  author = {Turakhia, Dishita and Ludgin, David and Mueller, Stefanie and Desportes, Kayla},
  title = {Understanding the Educators' Practices in Makerspaces for the Design of Education Tools},
  journal = {Educational Technology Research and Development},
  volume = {72},
  number = {1},
  pages = {329--358},
  year = {2024},
  doi = {10.1007/s11423-023-10305-1}
}

@inproceedings{turakhia2023reflectiveAR,
  author = {Turakhia, Dishita G. and Jiang, Peiling and Mueller, Stefanie},
  title = {The Reflective Make-{AR} In-Action: Using Augmented Reality for Reflection-based Learning of Makerskills},
  booktitle = {Extended Abstracts of the 2023 CHI Conference on Human Factors in Computing Systems},
  year = {2023},
  publisher = {ACM},
  doi = {10.1145/3544549.3585850}
}

%


\end{document}